\documentclass[12pt]{article}
\usepackage{amsmath,amssymb}
\usepackage{graphicx}
\newcommand{\eqdef}{\stackrel{\text{def}}{=}}

\newcommand{\n}{\nonumber\\}
\newcommand{\bm}{\boldsymbol}
\newcommand{\ignore}[1]{}
\numberwithin{equation}{section}
\newcommand{\Romannumeral}[1]{\uppercase\expandafter{\romannumeral#1}}
\newcommand{\I}{\text{\Romannumeral{1}}}
\newcommand{\II}{\text{\Romannumeral{2}}}

\allowdisplaybreaks[4]

\begin{document}

\baselineskip=20pt

\newcommand{\preprint}{
\vspace*{-20mm}
   \begin{flushright}\normalsize \sf
    DPSU-24-1\\
  \end{flushright}}
\newcommand{\Title}[1]{{\baselineskip=26pt
  \begin{center} \Large \bf #1 \\ \ \\ \end{center}}}
\newcommand{\Author}{\begin{center}
  \large \bf Satoru Odake \end{center}}
\newcommand{\Address}{\begin{center}
     Faculty of Science, Shinshu University,
     Matsumoto 390-8621, Japan
   \end{center}}
\newcommand{\Accepted}[1]{\begin{center}
  {\large \sf #1}\\ \vspace{1mm}{\small \sf Accepted for Publication}
  \end{center}}

\preprint
\thispagestyle{empty}

\Title{Type $\II$ Multi-indexed Little $q$-Jacobi and\\
Little $q$-Laguerre Polynomials}

\Author

\Address
\vspace{1cm}

\begin{abstract}
For the isospectral Darboux transformations of the discrete quantum mechanics
with real shifts, there are two methods: type $\I$ and type $\II$ constructions.
Based on the type $\I$ construction, the type $\I$ multi-indexed little
$q$-Jacobi and little $q$-Laguerre orthogonal polynomials were presented in
J.\,Phys.\,{\bf A50} (2017) 165204. 
Based on the type $\II$ construction, we present the type $\II$ multi-indexed
little $q$-Jacobi and little $q$-Laguerre orthogonal polynomials.
\end{abstract}

\section{Introduction}
\label{sec:intro}

The new type of orthogonal polynomials
 -- exceptional and multi-indexed polynomials --
\cite{gkm08}--\cite{idQMcH}
have the characteristic that they form a complete set of orthogonal basis
in spite of the missing degrees, by which the restrictions of Bochner's
theorem \cite{ismail} are avoided.
They are constructed based on the polynomials in the Askey-scheme of
hypergeometric orthogonal polynomials \cite{kls}, which satisfy
second order differential or difference equations.
To study such orthogonal polynomials, the quantum mechanical formulation is
very useful. We consider ordinary quantum mechanics (oQM) and two kinds of
discrete quantum mechanics (dQM), dQM with pure imaginary shifts (idQM) and
dQM with real shifts (rdQM) \cite{os24}. The Schr\"odinger equation for oQM
is a differential equation and that for dQM is a difference equation.
The coordinate $x$ for oQM and idQM is continuous and that for rdQM is discrete.

The multi-indexed orthogonal polynomials are systematically constructed by the
multi-step Darboux transformations for quantum mechanical systems.
When the wavefunctions of the virtual states are used as seed solutions,
the multi-step Darboux transformations give the case-(1) multi-indexed
orthogonal polynomials.
Here, the case-(1) is the case that the set of missing degrees of the
multi-indexed polynomials is $\{0,1,\ldots,\ell-1\}$, and the case-(2) is
otherwise.
The quantum mechanical systems associated to the case-(1) multi-indexed
orthogonal polynomials have shape invariance
\cite{os25}--\cite{idQMcH}.

For oQM (Jacobi and Laguerre \cite{os25}) and idQM (Askey-Wilson and Wilson
\cite{os27}, continuous Hahn \cite{idQMcH}) cases,
there are two types of virtual states, type $\I$ and type $\II$.
For rdQM (finite: $q$-Racah and Racah \cite{os26}, semi-infinite: Meixner,
little $q$-Jacobi and little $q$-Laguerre \cite{os35}) cases, we considered
one type of virtual state.
In the first manuscript of \cite{os26} (arXiv:1203.5868v1), we considered
two types of virtual states (type $\I$ and $\II$) for Racah and $q$-Racah
cases. But the multi-step Darboux transformations with these two types of
virtual states as seed solutions give essentially the same multi-indexed
polynomials, because these rdQM systems are finite systems.
So we discussed only one type of virtual state in \cite{os26}.
However, this situation may be different for semi-infinite systems.
The purpose of this paper is to consider two types of virtual states for the
semi-infinite rdQM systems and to obtain the type $\II$ multi-indexed little
$q$-Jacobi and little $q$-Laguerre polynomials.

This paper is organized as follows.
In section \ref{sec:DTrdQM} the finite and semi-infinite rdQM systems are
recapitulated and the multi-step Darboux transformations are discussed.
There are two methods, the type $\I$ and type $\II$ constructions.
Section \ref{sec:mioplqJ} is the main part of the paper.
Based on the type $\II$ construction, we obtain the case-(1) type $\II$
multi-indexed little $q$-Jacobi polynomials.
Similarly, the case-(1) type $\II$ multi-indexed little $q$-Laguerre
polynomials are obtained in section \ref{sec:mioplqL}.
Section \ref{sec:summary} is for a summary and comments.

\section{Darboux Transformations for rdQM}
\label{sec:DTrdQM}

In this section we recapitulate the isospectral Darboux transformations for
rdQM systems \cite{os26}.
The first manuscript of \cite{os26} is arXiv:1203.5868v1 and we will cite it
as \cite{os26}(v1).

\subsection{rdQM systems}
\label{sec:rdQM}

The Hamiltonian $\mathcal{H}$ of a finite rdQM system is a real symmetric
matrix, and we consider a tri-diagonal one \cite{os12},
\begin{align}
  \mathcal{H}&=(\mathcal{H}_{x,y})_{x,y\in\{0,1,\ldots,N\}},\\
  \mathcal{H}_{x,y}&=-\sqrt{B(x)D(x+1)}\,\delta_{x+1,y}
  -\sqrt{B(x-1)D(x)}\,\delta_{x-1,y}+\bigl(B(x)+D(x)\bigr)\delta_{x,y},
\end{align}
where the potential functions $B(x)$ and $D(x)$ are positive but vanish at
the boundary,
\begin{align}
  &B(x)>0\ \ (x=0,1,\ldots,N-1),\ \ B(N)=0,\n
  &D(x)>0\ \ (x=1,2,\ldots,N),\ \ D(0)=0.
  \label{BDcond}
\end{align}
We write this $\mathcal{H}$ as
\begin{align}
  \mathcal{H}&=-\sqrt{B(x)D(x+1)}\,e^{\partial}
  -\sqrt{B(x-1)D(x)}\,e^{-\partial}+B(x)+D(x)\n
  &=-\sqrt{B(x)}\,e^{\partial}\sqrt{D(x)}
  -\sqrt{D(x)}\,e^{-\partial}\sqrt{B(x)}+B(x)+D(x),
  \label{H}
\end{align}
where $e^{\pm\partial}$ is a matrix whose $(x,y)$-element is
$\delta_{x\pm 1,y}$, and $F(x)$ means a diagonal matrix
$F(x)=\text{diag}(F(0),F(1),\ldots,F(N))$.
The Schr\"odinger equation of rdQM system is a matrix eigenvalue problem,
\begin{equation}
  \mathcal{H}\phi_n(x)=\mathcal{E}_n\phi_n(x)
  \ \ (n=0,1,\ldots,N),\quad
  0=\mathcal{E}_0<\mathcal{E}_1<\cdots<\mathcal{E}_N,
  \label{Scheq}
\end{equation}
where the eigenstate vector (eigenvector) is
$\phi_n=(\phi_n(x))_{x\in\{0,1,\ldots,N\}}$
and the product of $\mathcal{H}$ and $\phi_n$ is given by
$\mathcal{H}\phi_n(x)\eqdef\sum_{y=0}^N\mathcal{H}_{x,y}\phi_n(y)$.
The constant term of $\mathcal{H}$ is chosen so that $\mathcal{E}_0=0$.
We remark that the matrix notation $\mathcal{H}\phi_n(x)$ represent
\begin{align*}
  \text{(a)}
  &:\ \mathcal{H}_{0,0}\phi_n(0)+\mathcal{H}_{0,1}\phi_n(1)\ \ (x=0),\\
  \text{(b)}
  &:\ \mathcal{H}_{x,x-1}\phi_n(x-1)+\mathcal{H}_{x,x}\phi_n(x)
  +\mathcal{H}_{x,x+1}\phi_n(x+1)\ \ (1\leq x\leq N-1),\\
  \text{(c)}
  &:\ \mathcal{H}_{N,N-1}\phi_n(N-1)+\mathcal{H}_{N,N}\phi_n(N)\ \ (x=N),
\end{align*}
and the expressions (a) and (c) can be regarded as given by (b)
because the boundary conditions $D(0)=0$ and $B(N)=0$ \eqref{BDcond} give
``$\,\mathcal{H}_{0,-1}$''$=0$ and ``$\,\mathcal{H}_{N,N+1}$''$=0$.
The inner product of two state vectors $f(x)$ and $g(x)$ is
$(f,g)=\sum_{x=0}^Nf(x)g(x)$ and the norm of $f(x)$ is $\|f\|=\sqrt{(f,f)}$.
The orthogonality relations for $\phi_n(x)$ are
\begin{equation}
  \quad(\phi_n,\phi_m)=\frac{1}{d_n^2}\delta_{nm}
  \ \ (n,m=0,1,\ldots,N),
  \label{orthophin}
\end{equation}
where $d_n$ are constants.

The Hamiltonian \eqref{H} can be expressed in factorized form in two ways
(type-(\romannumeral1) and type-(\romannumeral2) factorizations)
\cite{os26}(v1) :
\begin{align}
  \mathcal{H}&=\mathcal{A}^{\dagger}\mathcal{A}
  =\mathcal{A}^{(\text{\romannumeral2})\,\dagger}
  \mathcal{A}^{(\text{\romannumeral2})},\\
  \text{type-(\romannumeral1)}
  &:\ \ \mathcal{A}=\sqrt{B(x)}-e^{\partial}\sqrt{D(x)},\quad
  \mathcal{A}^{\dagger}=\sqrt{B(x)}-\sqrt{D(x)}\,e^{-\partial},
  \label{type(i)}\\
  \text{type-(\romannumeral2)}
  &:\ \ \mathcal{A}^{(\text{\romannumeral2})}
  =\sqrt{D(x)}-e^{-\partial}\sqrt{B(x)},\quad
  \mathcal{A}^{(\text{\romannumeral2})\,\dagger}
  =\sqrt{D(x)}-\sqrt{B(x)}\,e^{\partial}.
  \label{type(ii)}
\end{align}
We remark that since the factor $D(0)$ does not appear in $\mathcal{A}$ and
$\mathcal{A}^{\dagger}$, the boundary condition ``$D(0)=0$'' is automatically
satisfied in the type-(\romannumeral1) factorization ($B(N)=0$ is needed).
Similarly, since the factor $B(N)$ does not appear in
$\mathcal{A}^{(\text{\romannumeral2})}$ and
$\mathcal{A}^{(\text{\romannumeral2})\,\dagger}$, the boundary condition
``$B(N)=0$'' is automatically satisfied in the type-(\romannumeral2)
factorization ($D(0)=0$ is needed).

For the rdQM systems associated to the orthogonal polynomials in the
Askey scheme, the eigenstate $\phi_n(x)$ ($n\in\mathbb{Z}_{\geq 0}$) has the
following form \cite{os12},
\begin{equation}
  \phi_n(x)=\phi_0(x)\check{P}_n(x),\quad
  \check{P}_n(x)\eqdef P_n\bigl(\eta(x)\bigr).
\end{equation}
Here $\eta(x)$ is the sinusoidal coordinate \cite{os12} and $P_n(\eta)$ is a
polynomial of degree $n$ in $\eta$.
We choose the normalization $\check{P}_n(0)=1$ with $\eta(0)=0$, and set
$\check{P}_n(x)=0$ ($n\in\mathbb{Z}_{<0}$).
The ground state $\phi_0(x)$ is characterized by $\mathcal{A}\phi_0(x)=0$
(or $\mathcal{A}^{(\text{\romannumeral2})}\phi_0(x)=0$) and its explicit form
is
\begin{equation}
  \phi_0(x)=\sqrt{\prod_{y=0}^{x-1}\frac{B(y)}{D(y+1)}},
  \label{phi0}
\end{equation}
with the normalization $\phi_0(0)=1$ (convention: $\prod_{k=n}^{n-1}*=1$).
In the concrete examples, $B(x)$ and $D(x)$ are rational functions of $x$ or
$q^x$, and the defining range of $\phi_0(x)$ can be extended to $x\in\mathbb{R}$
(e.g., $(a;q)_x$ is expressed as $(a;q)_x=(a;q)_{\infty}/(aq^x;q)_{\infty}$,
which is defined for $x\in\mathbb{R}$).
The similarity transformed Hamiltonian $\widetilde{\mathcal{H}}$ in terms of
the ground state $\phi_0(x)$ is
\begin{equation}
  \widetilde{\mathcal{H}}\eqdef\phi_0(x)^{-1}\circ\mathcal{H}\circ\phi_0(x)
  =B(x)(1-e^{\partial})+D(x)(1-e^{-\partial}),
  \label{Ht}
\end{equation}
and \eqref{Scheq} becomes
\begin{equation}
  \widetilde{\mathcal{H}}\check{P}_n(x)=\mathcal{E}_n\check{P}_n(x).
\end{equation}

The semi-infinite rdQM systems, whose coordinate $x$ takes values in
$\mathbb{Z}_{\geq 0}$, can be obtained by taking $N\to\infty$ limit.

\subsection{Multi-step Darboux Transformations}
\label{sec:DT}

The property of the Darboux transformations depends on the choice of seed
solutions. We consider the virtual states as seed `solutions', which give
isospectral deformations.
There are two types of virtual states, the type I and type II \cite{os26}(v1).
For the construction of the multi-indexed ($q$-)Racah polynomials, these two
types of virtual states give essentially the same polynomials.
So the type II construction is omitted in \cite{os26}.

\noindent{\bf Remark}:
We comment on \cite{os26}(v1).
Instead of $\mathfrak{t}^{(\text{ex})}$ (in the last paragraph of \S\,3.4
of \cite{os26}(v1)), we should consider
$\mathfrak{t}^{(\text{ex}')}$:
\begin{equation}
  \mathfrak{t}^{(\text{ex}')}(\bm{\lambda})
  \eqdef(\lambda_1,\lambda_1+\lambda_3-\lambda_4,\lambda_1+\lambda_2-\lambda_4,
  2\lambda_1-\lambda_4),
\end{equation}
with $\lambda_1=-N$.
Then we have
\begin{equation}
  B^{\prime\,\II}(x;\bm{\lambda})
  =D^{\prime\,\I}\bigl(N-x;
  \mathfrak{t}^{(\text{ex}')}(\bm{\lambda})\bigr),\quad
  D^{\prime\,\II}(x;\bm{\lambda})
  =B^{\prime\,\I}\bigl(N-x;
  \mathfrak{t}^{(\text{ex}')}(\bm{\lambda})\bigr),
\end{equation}
and the type $\I$ and $\II$ multi-indexed ($q$-)Racah polynomials are related:
\begin{equation}
  \check{P}^{\II}_{\mathcal{D},n}(x;\bm{\lambda})
  =\check{P}^{\I}_{\mathcal{D},n}\bigl(N-x;\mathfrak{t}^{(\text{ex}')}
  (\bm{\lambda})\bigr).
\end{equation}

\medskip

Let us consider a finite rdQM system. We assume the existence of two rational
functions $B'(x)$ and $D'(x)$ of $x$ or $q^x$ satisfying
\begin{alignat}{2}
  B(x)D(x+1)&=\alpha^2B'(x)D'(x+1),&\alpha&>0,
  \label{BD=B'D'}\\
  B(x)+D(x)&=\alpha\bigl(B'(x)+D'(x)\bigr)+\alpha',&\qquad\alpha'&<0,
  \label{BD=B'D'2}
\end{alignat}
where $\alpha$ and $\alpha'$ are constants.
We impose on them the following conditions:
\begin{align}
  \text{type $\I$}:\ \ &B'(x)>0\ \ (x=0,1,\ldots,N+L-1),\n
  &D'(x)>0\ \ (x=1,2,\ldots,N),\ \ D'(0)=D'(N+1)=0,
  \label{B'>0,..}\\
  \text{type $\II$}:\ \ &D'(x)>0\ \ (x=-L+1,\ldots,-1,0,1,\ldots,N),\n
  &B'(x)>0\ \ (x=0,1,\ldots,N-1),\ \ B'(N)=B'(-1)=0,
  \label{B'>0,..II}
\end{align}
where $L$ is a certain positive integer to be specified later.
The function $\tilde{\phi}_0(x)$ is defined by
\begin{equation}
  \tilde{\phi}_0(x)\eqdef\sqrt{\prod_{y=0}^{x-1}\frac{B'(y)}{D'(y+1)}}
  \ \ (x=0,1,\ldots,N),
  \label{phit0}
\end{equation}
with the normalization $\tilde{\phi}_0(0)=1$.
Like $\phi_0(x)$, the defining range of $\tilde{\phi}_0(x)$ can be extended
to $x\in\mathbb{R}$.
We define a function $\nu(x)$,
\begin{equation}
  \nu(x)\eqdef\frac{\phi_0(x)}{\tilde{\phi}_0(x)}
  =\prod_{y=0}^{x-1}\frac{B(y)}{\alpha B'(y)}
  =\prod_{y=0}^{x-1}\frac{\alpha D'(y+1)}{D(y+1)}
  \ \ (x=0,1,\ldots,N),
  \label{nu}
\end{equation}
whose defining range can also be extended to $x\in\mathbb{R}$.
Note that $\nu^{\I}(x)=0$ for $x\in\mathbb{Z}_{\geq N+1}$
and $\nu^{\II}(x)=0$ for $x\in\mathbb{Z}_{\leq-1}$.
The relations \eqref{BD=B'D'}--\eqref{BD=B'D'2} imply the following relation
between two Hamiltonians:
\begin{align}
  \mathcal{H}&=\alpha\mathcal{H}'+\alpha',
  \label{H=aH'+a'}\\
  \mathcal{H}'&\eqdef-\sqrt{B'(x)}\,e^{\partial}\sqrt{D'(x)}
  -\sqrt{D'(x)}\,e^{-\partial}\sqrt{B'(x)}+B'(x)+D'(x).
  \label{H'}
\end{align}

We assume the existence of virtual state vectors $\tilde{\phi}_{\text{v}}(x)$
($\text{v}\in\mathcal{V}$),
\begin{equation}
  \tilde{\phi}_{\text{v}}(x)\eqdef\tilde{\phi}_0(x)\check{\xi}_{\text{v}}(x),
  \quad\check{\xi}_{\text{v}}(x)\eqdef\xi_{\text{v}}\bigl(\eta(x)\bigr).
  \label{phitv}
\end{equation}
Here $\mathcal{V}$ is the index set of the virtual state vectors, and
the virtual state polynomial $\xi_{\text{v}}(\eta)$ is a polynomial of degree
$\text{v}$ in $\eta$ satisfying the difference equation (for $x\in\mathbb{R}$)
\begin{equation}
  B'(x)\bigl(\check{\xi}_{\text{v}}(x)-\check{\xi}_{\text{v}}(x+1)\bigr)
  +D'(x)\bigl(\check{\xi}_{\text{v}}(x)-\check{\xi}_{\text{v}}(x-1)\bigr)
  =\mathcal{E}'_{\text{v}}\,\check{\xi}_{\text{v}}(x),
  \label{xieq}
\end{equation}
where $\mathcal{E}'_{\text{v}}$ is a constant.
We impose on $\mathcal{E}'_{\text{v}}$ and $\check{\xi}_{\text{v}}(x)$
the following conditions:
\begin{align}
  \tilde{\mathcal{E}}_{\text{v}}\eqdef\alpha\mathcal{E}'_{\text{v}}+\alpha',
  &\quad\tilde{\mathcal{E}}_{\text{v}}<0,
  \label{Etv<0}\\
  \text{type $\I$}:\ \ \check{\xi}_{\text{v}}(x)&>0\ \ (x=0,1,\ldots,N,N+1),
  \label{xiI>0,..}\\
  \text{type $\II$}:\ \ \check{\xi}_{\text{v}}(x)&>0\ \ (x=-1,0,1,\ldots,N).
  \label{xiII>0,..}
\end{align}
Since the matrix $\mathcal{H}$ is a positive semi-definite real symmetric
matrix, $\tilde{\mathcal{E}}_{\text{v}}$ being negative is a sufficient
condition for the virtual state $\tilde{\phi}_{\text{v}}(x)$ to not be a true
eigenvector, see \eqref{SchphitvI}--\eqref{SchphitvII}.
We choose the normalization $\check{\xi}_{\text{v}}(0)=1$ for type $\I$ and
$\check{\xi}_{\text{v}}(-1)=1$ for type $\II$ (which is different from
\cite{os26}(v1)).
Relations \eqref{H=aH'+a'}, \eqref{phitv} and \eqref{xieq} imply that
virtual state vectors $\tilde{\phi}_{\text{v}}(x)$ are polynomial `solutions'
of the Schr\"odinger equation except for one end-point:
\begin{alignat}{2}
  \text{type $\I$}:\ \ &\mathcal{H}\tilde{\phi}_{\text{v}}(x)
  =\tilde{\mathcal{E}}_{\text{v}}\tilde{\phi}_{\text{v}}(x)
  \ \ (x=0,1,\ldots,N-1),&\quad
  \mathcal{H}\tilde{\phi}_{\text{v}}(x)
  &\neq\tilde{\mathcal{E}}_{\text{v}}\tilde{\phi}_{\text{v}}(x)
  \ \ (x=N),
  \label{SchphitvI}\\
  \text{type $\II$}:\ \ &\mathcal{H}\tilde{\phi}_{\text{v}}(x)
  =\tilde{\mathcal{E}}_{\text{v}}\tilde{\phi}_{\text{v}}(x)
  \ \ (x=1,2,\ldots,N),&\quad
  \mathcal{H}\tilde{\phi}_{\text{v}}(x)
  &\neq\tilde{\mathcal{E}}_{\text{v}}\tilde{\phi}_{\text{v}}(x)
  \ \ (x=0),
  \label{SchphitvII}
\end{alignat}
due to $D'(0)=0$ and $B'(N)>0$ for type $\I$ \eqref{B'>0,..}, and
$D'(0)>0$ and $B'(N)=0$ for type $\II$ \eqref{B'>0,..II}.

For the Darboux transformation with the type $\I$ virtual state vector
$\tilde{\phi}^{\I}_{\text{v}}(x)$ (type $\II$ virtual state vector
$\tilde{\phi}^{\II}_{\text{v}}(x)$) as a seed solution, the
type-(\romannumeral1) factorization \eqref{type(i)} (type-(\romannumeral2)
factorization \eqref{type(ii)}) is used, respectively:
\begin{alignat*}{2}
  \text{type $\I$}:&
  \ \begin{cases}
  \mathcal{H}=\hat{\mathcal{A}}_{\text{v}}^{\dagger}
  \hat{\mathcal{A}}_{\text{v}}+\tilde{\mathcal{E}}_{\text{v}}^{\I}\\
  \hat{\mathcal{A}}_{\text{v}}\tilde{\phi}_{\text{v}}^{\I}(x)=0
  \ \ (x=0,1,\ldots,N-1)
  \end{cases}
  &\!\!\Rightarrow\ &\begin{cases}
  \mathcal{H}^{\text{new}}\eqdef\hat{\mathcal{A}}_{\text{v}}
  \hat{\mathcal{A}}_{\text{v}}^{\dagger}+\tilde{\mathcal{E}}_{\text{v}}^{\I}\\
  \phi^{\text{new}}_n(x)\eqdef\hat{\mathcal{A}}_{\text{v}}\phi_n(x)\,,
  \end{cases}\\
  \text{type $\II$}:&
  \ \begin{cases}
  \mathcal{H}=\hat{\mathcal{A}}_{\text{v}}^{(\text{\romannumeral2})\,\dagger}
  \hat{\mathcal{A}}_{\text{v}}^{(\text{\romannumeral2})}
  +\tilde{\mathcal{E}}_{\text{v}}^{\II}\\
  \hat{\mathcal{A}}_{\text{v}}^{(\text{\romannumeral2})}
  \tilde{\phi}_{\text{v}}^{\II}(x)=0
  \ \ (x=1,2,\ldots,N)    
  \end{cases}
  &\!\!\Rightarrow\ &\begin{cases}
  \mathcal{H}^{\text{new}}\eqdef
  \hat{\mathcal{A}}_{\text{v}}^{(\text{\romannumeral2})}
  \hat{\mathcal{A}}_{\text{v}}^{(\text{\romannumeral2})\,\dagger}
  +\tilde{\mathcal{E}}_{\text{v}}^{\II}\\
  \phi^{\text{new}}_n(x)\eqdef
  \hat{\mathcal{A}}_{\text{v}}^{(\text{\romannumeral2})}\phi_n(x)\,,
  \end{cases}
\end{alignat*}
and new virtual state vectors are
$\tilde{\phi}^{\I\,\text{new}}_{\text{v}'}(x)\eqdef
\hat{\mathcal{A}}_{\text{v}}\tilde{\phi}^{\I}_{\text{v}'}(x)
+(\text{correction term at $x=N$})$ and
$\tilde{\phi}^{\II\,\text{new}}_{\text{v}'}(x)\eqdef
\hat{\mathcal{A}}_{\text{v}}^{(\text{\romannumeral2})}
\tilde{\phi}^{\II}_{\text{v}'}(x)+(\text{correction term at $x=0$})$,
see \cite{os26}(v1) for explicit formulas.
Here we explain why the type-(\romannumeral1) (type-(\romannumeral2))
factorization is used for the type $\I$ (type $\II$) virtual state
vector, respectively.
The matrices $\hat{\mathcal{A}}_{\text{v}}$ and
$\hat{\mathcal{A}}_{\text{v}}^{(\text{\romannumeral2})}$ are given by
$\hat{\mathcal{A}}_{\text{v}}\eqdef
\sqrt{\hat{B}_{\text{v}}(x)}-e^{\partial}\sqrt{\hat{D}_{\text{v}}(x)}$ and
$\hat{\mathcal{A}}_{\text{v}}^{(\text{\romannumeral2})}\eqdef
\sqrt{\hat{D}_{\text{v}}(x)}-e^{-\partial}\sqrt{\hat{B}_{\text{v}}(x)}$ with
$\hat{B}_{\text{v}}(x)\eqdef\alpha B'(x)
\frac{\check{\xi}_{\text{v}}(x+1)}{\check{\xi}_{\text{v}}(x)}$ and
$\hat{D}_{\text{v}}(x)\eqdef\alpha D'(x)
\frac{\check{\xi}_{\text{v}}(x-1)}{\check{\xi}_{\text{v}}(x)}$, and they
satisfy
$\hat{B}_{\text{v}}(x)\hat{D}_{\text{v}}(x+1)=B(x)D(x+1)$ and
$\hat{B}_{\text{v}}(x)+\hat{D}_{\text{v}}(x)+\tilde{\mathcal{E}}_{\text{v}}
=B(x)+D(x)$.
Since $(0,0)$-element of
$\hat{\mathcal{A}}_{\text{v}}^{\dagger}
\hat{\mathcal{A}}_{\text{v}}+\tilde{\mathcal{E}}_{\text{v}}$ is
$\hat{B}_{\text{v}}(0)+\tilde{\mathcal{E}}_{\text{v}}$,
we need the condition $\hat{D}_{\text{v}}(0)=0$ for type-(\romannumeral1)
factorization.
Similary, since $(N,N)$-element of
$\hat{\mathcal{A}}_{\text{v}}^{(\text{\romannumeral2})\,\dagger}
\hat{\mathcal{A}}_{\text{v}}^{(\text{\romannumeral2})}
+\tilde{\mathcal{E}}_{\text{v}}$ is
$\hat{D}_{\text{v}}(N)+\tilde{\mathcal{E}}_{\text{v}}$,
we need the condition $\hat{B}_{\text{v}}(N)=0$ for type-(\romannumeral2)
factorization.
Be careful not to confuse
($\mathcal{A}$, $\mathcal{A}^{(\text{\romannumeral2})}$) and
($\hat{\mathcal{A}}_{\text{v}}$,
$\hat{\mathcal{A}}_{\text{v}}^{(\text{\romannumeral2})}$).
We call the multi-step Darboux transformations with only the type $\I$
(type $\II$) virtual state vectors as seed solutions type $\I$ (type $\II$)
construction.
In the multi-step Darboux transformations, various quantities are neatly
expressed in terms of Casoratians: $W_{\text{C}}$ (for type $\I$) and
$W_{\text{C}}^{(-)}$ (for type $\II$) \cite{os26}(v1).
The Casorati determinants of a set of $n$ functions $\{f_j(x)\}$ are defined by
\begin{align}
  \text{W}_{\text{C}}[f_1,f_2,\ldots,f_n](x)
  &\eqdef\det\Bigl(f_k(x+j-1)\Bigr)_{1\leq j,k\leq n},
  \label{WC}\\
  \text{W}^{(-)}_{\text{C}}[f_1,f_2,\ldots,f_n](x)
  &\eqdef\det\Bigl(f_k(x-j+1)\Bigr)_{1\leq j,k\leq n}\n
  &=(-1)^{\binom{n}{2}}\text{W}_{\text{C}}[f_1,f_2,\ldots,f_n](x-n+1),
  \label{WC-}
\end{align}
(for $n=0$, we set $\text{W}_{\text{C}}[\cdot](x)=
\text{W}^{(-)}_{\text{C}}[\cdot](x)=1$).
The auxiliary functions $\varphi(x)$, $\varphi_M(x)$ and $\varphi_M^{(-)}(x)$
($M\in\mathbb{Z}_{\geq 0}$) are defined by \cite{os12,os22} \cite{os26}(v1)
\begin{align}
  \varphi(x)&\eqdef\frac{\eta(x+1)-\eta(x)}{\eta(1)},
  \label{varphidef}\\
  \varphi_M(x)&\eqdef\prod_{1\leq j<k\leq M}
  \frac{\eta(x+k-1)-\eta(x+j-1)}{\eta(k-j)},
  \label{varphiMdef}\\
  \varphi^{(-)}_M(x)&\eqdef\prod_{1\leq j<k\leq M}
  \frac{\eta(x-j+1)-\eta(x-k+1)}{\eta(k-j)}
  =\varphi_M(x-M+1),
  \label{varphiMdefII}
\end{align}
and $\varphi_0(x)=\varphi_1(x)=\varphi^{(-)}_0(x)=\varphi^{(-)}_1(x)=1$.

Let us consider the $M$-step Darboux transformations with virtual state vectors
$\tilde{\phi}_{\text{v}}(x)$ ($\text{v}\in\mathcal{D}$) as seed solutions.
Here $\mathcal{D}$ is
\begin{equation}
  \mathcal{D}=\{d_1,d_2,\ldots,d_M\}
  \ \ (1\leq d_1<d_2<\cdots<d_M\ ;\ d_j\in\mathcal{V}),
  \label{setD}
\end{equation}
and we assume $|\mathcal{V}|\geq M$ and $L\geq M$.
Although this notation $d_j$ conflicts with the notation of the normalization
constant $d_n$ in \eqref{orthophin}, we think this does not cause any confusion
because the latter appears as $\frac{1}{d_n^2}\,\delta_{nm}$.
The Hamiltonian of the deformed system, the Schr\"odinger equation and
the orthogonality relations are given by \cite{os26}
\begin{align}
  &\mathcal{H}_{\mathcal{D}}
  =(\mathcal{H}_{\mathcal{D}\,x,y})_{x,y\in\{0,1,\ldots,N\}}\n
  &\phantom{\mathcal{H}_{\mathcal{D}}}
  \eqdef-\sqrt{B_{\mathcal{D}}(x)}\,e^{\partial}\sqrt{D_{\mathcal{D}}(x)}
  -\sqrt{D_{\mathcal{D}}(x)}\,e^{-\partial}\sqrt{B_{\mathcal{D}}(x)}
  +B_{\mathcal{D}}(x)+D_{\mathcal{D}}(x)
  \label{HD}\\
  &\phantom{\mathcal{H}_{\mathcal{D}}}
  =\mathcal{A}_{\mathcal{D}}^{\dagger}\mathcal{A}_{\mathcal{D}},\quad
  \mathcal{A}_{\mathcal{D}}\eqdef\sqrt{B_{\mathcal{D}}(x)}
  -e^{\partial}\sqrt{D_{\mathcal{D}}(x)},
  \ \ \mathcal{A}_{\mathcal{D}}^{\dagger}
  \eqdef\sqrt{B_{\mathcal{D}}(x)}-\sqrt{D_{\mathcal{D}}(x)}\,e^{-\partial},\\
  &\mathcal{H}_{\mathcal{D}}\phi_{\mathcal{D}\,n}(x)
  =\mathcal{E}_n\phi_{\mathcal{D}\,n}(x)
  \ \ (n=0,1,\ldots,N),
  \label{SchD}\\
  &(\phi_{\mathcal{D}\,n},\phi_{\mathcal{D}\,m})
  =\prod_{j=1}^M(\mathcal{E}_n-\tilde{\mathcal{E}}_{d_j})
  \cdot\frac{1}{d_n^2}\delta_{nm}
  \ \ (n,m=0,1,\ldots,N).
  \label{orthophiDn}
\end{align}
The semi-infinite systems are obtained by taking $N\to\infty$ limit.
We present explicit forms of $B_{\mathcal{D}}(x)$, $D_{\mathcal{D}}(x)$ and
$\phi_{\mathcal{D}\,n}(x)$ for the semi-infinite systems
in \S\,\ref{sec:DTI}--\ref{sec:DTII}.
Only the final results are given here. For intermediate steps, see
\cite{os26,os35}, \cite{os26}(v1).

\subsubsection{type $\I$ construction for semi-infinite systems}
\label{sec:DTI}

In the type $\I$ construction for semi-infinite systems, the virtual state
vectors $\tilde{\phi}_{\text{v}}(x)$ are now solutions of the Schr\"odinger
equation \eqref{SchphitvI} (with $N\to\infty$), but they have infinite norms.
Explicit forms of $B_{\mathcal{D}}(x)$, $D_{\mathcal{D}}(x)$ and
$\phi_{\mathcal{D}\,n}(x)$ are
\begin{align}
  B_{\mathcal{D}}(x)&\eqdef\alpha B'(x+M)
  \frac{\text{W}_{\text{C}}[\check{\xi}_{d_1},\ldots,\check{\xi}_{d_M}](x)}
  {\text{W}_{\text{C}}[\check{\xi}_{d_1},\ldots,\check{\xi}_{d_M}](x+1)}\,
  \frac{\text{W}_{\text{C}}[\check{\xi}_{d_1},\ldots,\check{\xi}_{d_M},\nu](x+1)}
  {\text{W}_{\text{C}}[\check{\xi}_{d_1},\ldots,\check{\xi}_{d_M},\nu](x)},
  \label{BDI}\\
  D_{\mathcal{D}}(x)&\eqdef\alpha D'(x)
  \frac{\text{W}_{\text{C}}[\check{\xi}_{d_1},\ldots,\check{\xi}_{d_M}](x+1)}
  {\text{W}_{\text{C}}[\check{\xi}_{d_1},\ldots,\check{\xi}_{d_M}](x)}\,
  \frac{\text{W}_{\text{C}}[\check{\xi}_{d_1},\ldots,\check{\xi}_{d_M},\nu](x-1)}
  {\text{W}_{\text{C}}[\check{\xi}_{d_1},\ldots,\check{\xi}_{d_M},\nu](x)},
  \label{DDI}\\
  \phi_{\mathcal{D}\,n}(x)&\eqdef
  \frac{(-1)^M\sqrt{\prod_{j=1}^M\alpha B'(x+j-1)}
  \,\tilde{\phi}_0(x)\,
  \text{W}_{\text{C}}[\check{\xi}_{d_1},\ldots,\check{\xi}_{d_M},
  \nu\check{P}_n](x)}
  {\sqrt{\text{W}_{\text{C}}[\check{\xi}_{d_1},\ldots,\check{\xi}_{d_M}](x)\,
  \text{W}_{\text{C}}[\check{\xi}_{d_1},\ldots,\check{\xi}_{d_M}](x+1)}}.
  \label{phiDnI}
\end{align}
The Casoratian
$\text{W}_{\text{C}}[\check{\xi}_{d_1},\ldots,\check{\xi}_{d_M}](x)$ has
definite sign for $x\in\mathbb{Z}_{\geq 0}$ (see \cite{os35} for proof).
The potential functions $B_{\mathcal{D}}(x)$ and $D_{\mathcal{D}}(x)$ are
positive: $B_{\mathcal{D}}(x)>0$ ($x\in\mathbb{Z}_{\geq 0}$),
$D_{\mathcal{D}}(x)>0$ ($x\in\mathbb{Z}_{\geq 1}$) and
$D_{\mathcal{D}}(0)=0$.

\subsubsection{type $\II$ construction for semi-infinite systems}
\label{sec:DTII}

In the type $\II$ construction for semi-infinite systems, the virtual state
vectors $\tilde{\phi}_{\text{v}}(x)$ satisfy the Schr\"odinger equation except
for the end-point $x=0$, \eqref{SchphitvII} (with $N\to\infty$).
Explicit forms of $B_{\mathcal{D}}(x)$, $D_{\mathcal{D}}(x)$ and
$\phi_{\mathcal{D}\,n}(x)$ are
\begin{align}
  B_{\mathcal{D}}(x)&\eqdef\alpha B'(x)
  \frac{\text{W}^{(-)}_{\text{C}}[\check{\xi}_{d_1},\ldots,
  \check{\xi}_{d_M}](x-1)}
  {\text{W}^{(-)}_{\text{C}}[\check{\xi}_{d_1},\ldots,\check{\xi}_{d_M}](x)}\,
  \frac{\text{W}^{(-)}_{\text{C}}[\check{\xi}_{d_1},\ldots,\check{\xi}_{d_M},
  \nu](x+1)}
  {\text{W}^{(-)}_{\text{C}}[\check{\xi}_{d_1},\ldots,\check{\xi}_{d_M},\nu](x)},
  \label{BDII}\\
  D_{\mathcal{D}}(x)&\eqdef\alpha D'(x-M)
  \frac{\text{W}^{(-)}_{\text{C}}[\check{\xi}_{d_1},\ldots,\check{\xi}_{d_M}](x)}
  {\text{W}^{(-)}_{\text{C}}[\check{\xi}_{d_1},\ldots,\check{\xi}_{d_M}](x-1)}\,
  \frac{\text{W}^{(-)}_{\text{C}}[\check{\xi}_{d_1},\ldots,\check{\xi}_{d_M},
  \nu](x-1)}
  {\text{W}^{(-)}_{\text{C}}[\check{\xi}_{d_1},\ldots,\check{\xi}_{d_M},\nu](x)},
  \label{DDII}\\
  \phi_{\mathcal{D}\,n}(x)&\eqdef
  \frac{(-1)^M\sqrt{\prod_{j=1}^M\alpha D'(x-j+1)}
  \,\tilde{\phi}_0(x)\,
  \text{W}^{(-)}_{\text{C}}[\check{\xi}_{d_1},\ldots,\check{\xi}_{d_M},
  \nu\check{P}_n](x)}
  {\sqrt{\text{W}^{(-)}_{\text{C}}[\check{\xi}_{d_1},\ldots,
  \check{\xi}_{d_M}](x)\,
  \text{W}^{(-)}_{\text{C}}[\check{\xi}_{d_1},\ldots,\check{\xi}_{d_M}](x-1)}}.
  \label{phiDnII}
\end{align}
The Casoratian
$\text{W}^{(-)}_{\text{C}}[\check{\xi}_{d_1},\ldots,\check{\xi}_{d_M}](x)$ has
definite sign for $x\in\mathbb{Z}_{\geq-1}$ (by similar proof as type $\I$).
The potential functions $B_{\mathcal{D}}(x)$ and $D_{\mathcal{D}}(x)$ are
positive: $B_{\mathcal{D}}(x)>0$ ($x\in\mathbb{Z}_{\geq 0}$),
$D_{\mathcal{D}}(x)>0$ ($x\in\mathbb{Z}_{\geq 1}$) and
$D_{\mathcal{D}}(0)=0$.
In \S\,\ref{sec:mioplqJ} and \S\,\ref{sec:mioplqL}, we will denote
$\phi_{\mathcal{D}\,n}(x)$ in \eqref{phiDnII} as
$\phi_{\mathcal{D}\,n}^{\text{gen}}(x)$.

\section{Multi-indexed Little $q$-Jacobi polynomials}
\label{sec:mioplqJ}

In this section we present the case-(1) multi-indexed little $q$-Jacobi
polynomials, especially type $\II$ polynomials.
Various quantities depend on a set of parameters
$\bm{\lambda}=(\lambda_1,\lambda_2,\ldots)$ and $q$ ($0<q<1$), and
$q^{\bm{\lambda}}$ stands for
$q^{(\lambda_1,\lambda_2,\ldots)}=(q^{\lambda_1},q^{\lambda_2},\ldots)$.
Their dependence is expressed as $f=f(\bm{\lambda})$ and
$f(x)=f(x;\bm{\lambda})$, but $q$-dependence is suppressed.

\subsection{Original system}
\label{sec:lqJ}

Let us present the basic data of little $q$-Jacobi rdQM system.
The standard little $q$-Jacobi polynomial
$p_n(q^x;a,b|q)$ $={}_2\phi_1\bigl(\genfrac{}{}{0pt}{}{q^{-n},\,abq^{n+1}}{aq}
\bigl|q\,;q^{x+1}\bigr)$ \cite{kls} does not satisfy our normalization
$\check{P}_n(0)=1$.
We change the parametrization slightly from the standard one,
$(a,b)^{\text{standard}}=(aq^{-1},bq^{-1})$ \cite{os39}.
The basic data are as follows \cite{os12,os34}:
\begin{align}
  &q^{\bm{\lambda}}=(a,b),\quad
  \bm{\delta}=(1,1),\quad \kappa=q^{-1},\quad 0<a<1,\quad b<1,\\
  &B(x;\bm{\lambda})=aq^{-1}(q^{-x}-b),\quad D(x)=q^{-x}-1,\\
  &\mathcal{E}_n(\bm{\lambda})=(q^{-n}-1)(1-abq^{n-1}),\quad
  \eta(x)=1-q^x,\quad\varphi(x)=q^x,\\
  &\check{P}_n(x;\bm{\lambda})
  ={}_3\phi_1\Bigl(
  \genfrac{}{}{0pt}{}{q^{-n},\,abq^{n-1},\,q^{-x}}{b}\Bigm|
  q\,;a^{-1}q^{x+1}\Bigr)
  =c'_n(\bm{\lambda})\,p_n\bigl(1-\eta(x);aq^{-1},bq^{-1}|q\bigr)\n
  &\phantom{\check{P}_n(x;\bm{\lambda})}
  =c'_n(\bm{\lambda})\,{}_2\phi_1\Bigl(
  \genfrac{}{}{0pt}{}{q^{-n},\,abq^{n-1}}{a}\Bigm|q\,;q^{x+1}\Bigr),\quad
  c'_n(\bm{\lambda})\eqdef(-a)^{-n}q^{-\binom{n}{2}}\frac{(a;q)_n}{(b;q)_n},\\
  &\check{P}_n(x;\bm{\lambda})=c_n(\bm{\lambda})\eta(x)^n
  +\text{lower degree terms},
  \quad c_n(\bm{\lambda})\eqdef(-a)^{-n}q^{-n(n-1)}
  \frac{(abq^{n-1};q)_n}{(b;q)_n},\\
  &\phi_0(x;\bm{\lambda})^2=\frac{(b;q)_x}{(q;q)_x}a^x
  =\frac{(b,q^x;q)_{\infty}}{(bq^x,q;q)_{\infty}}a^x,\quad
  \phi_0(x;\bm{\lambda})>0,\quad\phi_0(0;\bm{\lambda})=1,
  \label{lqJphi0}\\
  &d_n(\bm{\lambda})^2
  =\frac{(b,ab;q)_n\,a^nq^{n(n-1)}}{(a,q;q)_n}\,
  \frac{1-abq^{2n-1}}{1-abq^{n-1}}
  \times\frac{(a;q)_{\infty}}{(ab;q)_{\infty}},\quad
  d_n(\bm{\lambda})>0,
\end{align}
and $\check{P}_n(x;\bm{\lambda})$ satisfies
\begin{equation}
  \check{P}_n(0;\bm{\lambda})=1,\quad
  \check{P}_n(\infty;\bm{\lambda})\bigl(\,\eqdef
  \lim_{x\to\infty}\check{P}_n(x;\bm{\lambda})\bigr)
  =c'_n(\bm{\lambda}).
  \label{Pn(0)}
\end{equation}
Note that the most right hand side of $\phi_0(x;\bm{\lambda})^2$ \eqref{lqJphi0}
is defined for $x\in\mathbb{R}$.
This rdQM system is shape invariant,
\begin{equation}
  \mathcal{A}(\bm{\lambda})\mathcal{A}(\bm{\lambda})^{\dagger}
  =\kappa\mathcal{A}(\bm{\lambda}+\bm{\delta})^{\dagger}
  \mathcal{A}(\bm{\lambda}+\bm{\delta})+\mathcal{E}_1(\bm{\lambda}).
  \label{si}
\end{equation}
As a consequence of the shape invariance combined with the Crum's theorem
and the normalization, we obtain \cite{os24}
\begin{align}
  &\mathcal{A}(\bm{\lambda})\phi_n(x;\bm{\lambda})
  =\frac{\mathcal{E}_n(\bm{\lambda})}{\sqrt{B(0;\bm{\lambda})}}\,
  \phi_{n-1}(x;\bm{\lambda}+\bm{\delta})
  \ \ (n\in\mathbb{Z}_{\geq 0}),
  \label{Aphin=}\\
  &\mathcal{A}(\bm{\lambda})^{\dagger}\phi_{n-1}(x;\bm{\lambda}+\bm{\delta})
  =\sqrt{B(0;\bm{\lambda})}\,\phi_n(x;\bm{\lambda})
  \ \ (n\in\mathbb{Z}_{\geq 1}).
  \label{Adphin=}
\end{align}
These relations give the forward and backward shift relations:
\begin{align}
  &\mathcal{F}(\bm{\lambda})\check{P}_n(x;\bm{\lambda})
  =\mathcal{E}_n(\bm{\lambda})\check{P}_{n-1}(x;\bm{\lambda}+\bm{\delta})
  \ \ (n\in\mathbb{Z}_{\geq 0}),\\
  &\mathcal{B}(\bm{\lambda})\check{P}_{n-1}(x;\bm{\lambda}+\bm{\delta})
  =\check{P}_n(x;\bm{\lambda}),
  \ \ (n\in\mathbb{Z}_{\geq 1}),
\end{align}
where the forward and backward shift operators are
\begin{align}
  \mathcal{F}(\bm{\lambda})
  &\eqdef\sqrt{B(0;\bm{\lambda})}\,\phi_0(x;\bm{\lambda}+\bm{\delta})^{-1}
  \circ\mathcal{A}(\bm{\lambda})\circ\phi_0(x;\bm{\lambda})
  =B(0;\bm{\lambda})\varphi(x)^{-1}
  (1-e^{\partial}),\\
  \mathcal{B}(\bm{\lambda})
  &\eqdef\frac{1}{\sqrt{B(0;\bm{\lambda})}}\,
  \phi_0(x;\bm{\lambda})^{-1}\circ\mathcal{A}(\bm{\lambda})^{\dagger}
  \circ\phi_0(x;\bm{\lambda}+\bm{\delta})\n
  &=B(0;\bm{\lambda})^{-1}
  \bigl(B(x;\bm{\lambda})-D(x)e^{-\partial}\bigr)\varphi(x).
  \label{cBdef}
\end{align}
The similarity transformed Hamiltonian \eqref{Ht} is expressed as
$\widetilde{\mathcal{H}}(\bm{\lambda})=\mathcal{B}(\bm{\lambda})
\mathcal{F}(\bm{\lambda})$.
The auxiliary functions $\varphi_M(x)$ \eqref{varphiMdef} and
$\varphi^{(-)}_M(x)$ \eqref{varphiMdefII} become
\begin{equation}
  \varphi_M(x)=q^{\binom{M}{2}x+\binom{M}{3}},\quad
  \varphi^{(-)}_M(x)=q^{\binom{M}{2}x-\frac16M(M-1)(2M-1)}.
  \label{varphiM}
\end{equation}

\subsection{Type $\I$ polynomials}
\label{sec:lqJI}

The potential functions $B'(x)$ and $D'(x)$ and the virtual state polynomials
$\check{\xi}_{\text{v}}(x)$ are given by \cite{os35}
\begin{align}
  B^{\prime\,\I}(x;\bm{\lambda})&\eqdef
  B\bigl(x;\mathfrak{t}^{\I}(\bm{\lambda})\bigr),\quad
  D^{\prime\,\I}(x)\eqdef D(x),
  \label{B'I}\\
  \check{\xi}^{\I}_{\text{v}}(x;\bm{\lambda})&\eqdef
  \check{P}_{\text{v}}\bigl(x;\mathfrak{t}^{\I}(\bm{\lambda})\bigr),
  \label{xiI}
\end{align}
where the twist operation $\mathfrak{t}$ and the shift $\tilde{\bm{\delta}}$ are
(remark: $(a,b)^{\text{standard}}$ are used in \cite{os35})
\begin{align}
  &\mathfrak{t}^{\I}(\bm{\lambda})\eqdef(2-\lambda_1,\lambda_2),
  \ \ \text{namely}\ \ q^{\mathfrak{t}^{\I}(\bm{\lambda})}=(a^{-1}q^2,b),\\
  &\tilde{\bm{\delta}}^{\I}\eqdef(-1,1),\quad
  \mathfrak{t}^{\I}(\bm{\lambda})+u\bm{\delta}=
  \mathfrak{t}^{\I}(\bm{\lambda}+u\tilde{\bm{\delta}}^{\I})
  \ \ (\forall u\in\mathbb{R}),
\end{align}
with $\alpha^{\I}(\bm{\lambda})\eqdef aq^{-1}$ and
$\alpha^{\prime\,\I}(\bm{\lambda})\eqdef-(1-aq^{-1})(1-b)$.
The parameter range is $0<a<q^{1+d_M}$ and $b<1$.
Various formulas for the type $\I$ multi-indexed little $q$-Jacobi polynomials
are presented in \cite{os35}.

\subsection{Type $\II$ polynomials}
\label{sec:lqJII}

The twist operation $\mathfrak{t}$ and the shift $\tilde{\bm{\delta}}$ are
defined by
\begin{align}
  &\mathfrak{t}^{\II}(\bm{\lambda})\eqdef(\lambda_1,2-\lambda_2),
  \ \ \text{namely}\ \ q^{\mathfrak{t}^{\II}(\bm{\lambda})}=(a,b^{-1}q^2),
  \label{twistII}\\
  &\tilde{\bm{\delta}}^{\II}\eqdef(1,-1),\quad
  \mathfrak{t}^{\II}(\bm{\lambda})+u\bm{\delta}=
  \mathfrak{t}^{\II}(\bm{\lambda}+u\tilde{\bm{\delta}}^{\II})
  \ \ (\forall u\in\mathbb{R}).
  \label{tdeltaII}
\end{align}
Without using this twist operation, let us define the potential functions and
the virtual state polynomials.
The potential functions $B'(x)$ and $D'(x)$ are given by
\begin{equation}
  B^{\prime\,\II}(x;\bm{\lambda})\eqdef ab^{-1}q(q^{-x-1}-1),\quad
  D^{\prime\,\II}(x;\bm{\lambda})\eqdef b^{-1}q^{1-x}-1,
  \label{B'II}
\end{equation}
which satisfy the conditions \eqref{BD=B'D'}--\eqref{BD=B'D'2} with
\begin{equation}
  \alpha^{\II}(\bm{\lambda})\eqdef bq^{-1},\quad
  \alpha^{\prime\,\II}(\bm{\lambda})\eqdef-(1-a)(1-bq^{-1}).
  \label{alphaII}
\end{equation}
For \eqref{B'>0,..II} (with $N\to\infty$), we take $L=M$ and assume $0<b<q^M$.
The virtual state polynomial $\check{\xi}_{\text{v}}(x)$
($\text{v}\in\mathbb{Z}_{\geq 0}$) is given by
\begin{equation}
  \check{\xi}^{\II}_{\text{v}}(x;\bm{\lambda})\eqdef
  \tilde{c}^{\prime\,\II}_{\text{v}}(\bm{\lambda})\,{}_2\phi_1\Bigl(
  \genfrac{}{}{0pt}{}{q^{-\text{v}},\,ab^{-1}q^{\text{v}+1}}{a}
  \Bigm|q\,;bq^x\Bigr),\quad
  \tilde{c}^{\prime\,\II}_{\text{v}}(\bm{\lambda})\eqdef
  \frac{(a;q)_{\text{v}}}{(bq^{-\text{v}-1};q)_{\text{v}}},
  \label{xivII}
\end{equation}
which satisfies
\begin{equation}
  \check{\xi}^{\II}_{\text{v}}(-1;\bm{\lambda})=1,\quad
  \check{\xi}^{\II}_{\text{v}}(\infty;\bm{\lambda})\bigl(\,\eqdef
  \lim_{x\to\infty}\check{\xi}^{\II}_{\text{v}}(x;\bm{\lambda})\bigr)
  =\tilde{c}^{\prime\,\II}_{\text{v}}(\bm{\lambda}),
  \label{xiv(-1)}
\end{equation}
and
\begin{equation}
  \check{\xi}^{\II}_{\text{v}}(x;\bm{\lambda})
  =\tilde{c}^{\II}_{\text{v}}(\bm{\lambda})\eta(x)^{\text{v}}
  +\text{lower degree terms},\quad
  \tilde{c}^{\II}_{\text{v}}(\bm{\lambda})\eqdef
  b^{\text{v}}q^{-\binom{\text{v+1}}{2}}
  \frac{(ab^{-1}q^{\text{v}+1};q)_{\text{v}}}{(bq^{-\text{v}-1};q)_{\text{v}}}.
  \label{ctvII}
\end{equation}
As mentioned below \eqref{tdeltaII},
$\check{\xi}^{\II}_{\text{v}}(x;\bm{\lambda})$ is defined without using
$\mathfrak{t}^{\II}$, and we remark that
$\check{\xi}^{\II}_{\text{v}}(x;\bm{\lambda})\not\propto
\check{P}_{\text{v}}(x;\mathfrak{t}^{\II}(\bm{\lambda}))$
in contrast to the type $\I$ case \eqref{xiI}.
For simplicity of presentation, the superscript $\II$ is omitted in the
following.

The virtual state polynomial $\check{\xi}_{\text{v}}(x)$ satisfies the
difference equation (for $x\in\mathbb{R}$) \eqref{xieq} with
\begin{equation}
  \mathcal{E}'_{\text{v}}(\bm{\lambda})\eqdef
  \mathcal{E}_{\text{v}}\bigl(\mathfrak{t}(\bm{\lambda})\bigr)
  =(q^{-\text{v}}-1)(1-ab^{-1}q^{\text{v}+1}).
  \label{E'v}
\end{equation}
Proof:
Let us consider the following function $f_{\text{v}}(z)$
($\text{v}\in\mathbb{Z}_{\geq 0}$),
\begin{align*}
  f_{\text{v}}(z)=&\,{}_2\phi_1\Bigl(
  \genfrac{}{}{0pt}{}{q^{-\text{v}},\,ab^{-1}q^{\text{v}+1}}{a}
  \Bigm|q\,;bz\Bigr)=\sum_{k=0}^{\text{v}}a_kz^k,\quad
  a_k=\frac{(q^{-\text{v}},ab^{-1}q^{\text{v}+1};q)_k}{(a;q)_k}
  \frac{b^k}{(q;q)_k},\\
  &a_k=\frac{(1-q^{-\text{v}+k-1})(1-ab^{-1}q^{\text{v}+k})}{1-aq^{k-1}}
  \frac{b}{1-q^k}a_{k-1}\ \ (1\leq k\leq\text{v}).
\end{align*}
This $f_{\text{v}}(z)$ satisfies the $q$-difference equation
\begin{equation*}
  ab^{-1}(1-qz)\bigl(f_{\text{v}}(z)-f_{\text{v}}(qz)\bigr)
  +(b^{-1}q-z)\bigl(f_{\text{v}}(z)-f_{\text{v}}(q^{-1}z)\bigr)
  =(q^{-\text{v}}-1)(1-ab^{-1}q^{\text{v}+1})zf_{\text{v}}(z),
\end{equation*}
which is shown by comparing the coefficients of $z^k$ terms of both sides
($k=0$, $1\leq k\leq\text{v}$ and $k=\text{v}+1$).
By substituting $z=q^x$ and
$f_{\text{v}}(z)=\tilde{c}'_{\text{v}}{}^{-1}\check{\xi}_{\text{v}}(x)$ into
this $q$-difference equation, we obtain \eqref{xieq}.
\hfill\fbox{}

{}From \eqref{xiII>0,..} (with $N\to\infty$), the virtual state polynomials
should satisfy $\check{\xi}_{\text{v}}(x)>0$ ($x\in\mathbb{Z}_{\geq-1}$).
Let us check this condition.
Note that $\tilde{c}'_{\text{v}}>0$ for $a<1$ and $b<q^{\text{v}+1}$.
Since $\check{\xi}_{\text{v}}(-1)=1>0$, we consider $x\in\mathbb{Z}_{\geq 0}$.
By using the identity ((1.13.17) in \cite{kls})
\begin{equation*}
  {}_2\phi_1\Bigl(\genfrac{}{}{0pt}{}{q^{-n},\,b}{c}\Bigm|q;z\Bigr)
  =(bc^{-1}q^{-n}z;q)_n\,
  {}_3\phi_2\Bigl(\genfrac{}{}{0pt}{}{q^{-n},\,b^{-1}c,\,0}
  {c,\,b^{-1}cqz^{-1}}\Bigm|q;q\Bigr)\ \ (n\in\mathbb{Z}_{\geq 0}),
\end{equation*}
$\check{\xi}_{\text{v}}(x)$ is rewritten as
\begin{align}
  \check{\xi}_{\text{v}}(x)
  &=\tilde{c}'_{\text{v}}(q^{x+1};q)_{\text{v}}\,
  {}_3\phi_2\Bigl(\genfrac{}{}{0pt}{}{q^{-\text{v}},\,bq^{-\text{v}-1},\,0}
  {a,\,q^{-\text{v}-x}}\Bigm|q;q\Bigr)\n
  &=\tilde{c}'_{\text{v}}(q^{x+1};q)_{\text{v}}
  \sum_{k=0}^{\text{v}}
  \frac{(q^{-\text{v}},bq^{-\text{v}-1};q)_k}{(a,q^{-\text{v}-x};q)_k}
  \frac{q^k}{(q;q)_k}\n
  &=\tilde{c}'_{\text{v}}(q^{x+1};q)_{\text{v}}
  \sum_{k=0}^{\text{v}}
  \frac{(q^{\text{v}-k+1},bq^{-\text{v}-1};q)_k}{(a,q^{x+\text{v}-k+1};q)_k}
  \frac{q^{k(x+1)}}{(q;q)_k},
\end{align}
and each $k$-th term of the sum are positive for $a<1$ and $b<q^{\text{v}+1}$.
Therefore we obtain
$\check{\xi}_{\text{v}}(x)>0$ ($x\in\mathbb{Z}_{\geq -1}$) for $a<1$ and
$b<q^{\text{v}+1}$.

In the following, we assume the following parameter range:
\begin{equation}
  0<a<1,\quad 0<b<q^{1+d_M},
  \label{range}
\end{equation}
which will be extended in the last paragraph of this subsection.
The functions $\tilde{\phi}_0(x)$ ($>0$) \eqref{phit0} and $\nu(x)$ \eqref{nu}
become
\begin{align}
  \tilde{\phi}_0(x;\bm{\lambda})^2&=\frac{(q;q)_x}{(b;q)_x}a^x
  =\frac{(bq^x,q;q)_{\infty}}{(b,q^{x+1};q)_{\infty}}a^x,\\
  \nu(x;\bm{\lambda})&=\frac{(b;q)_x}{(q;q)_x}
  =\frac{(b,q^{x+1};q)_{\infty}}{(bq^x,q;q)_{\infty}},
\end{align}
and the virtual state energy $\tilde{\mathcal{E}}_{\text{v}}$ \eqref{Etv<0}
becomes
\begin{equation}
  \tilde{\mathcal{E}}_{\text{v}}(\bm{\lambda})
  =-(1-aq^{\text{v}})(1-bq^{-1-\text{v}}).
\end{equation}
The virtual state vectors $\tilde{\phi}_{\text{v}}(x)$ \eqref{phitv} satisfy
the Schr\"odinger equation except for the end-point $x=0$,
\eqref{SchphitvII} (with $N\to\infty$).

We define the denominator polynomial $\check{\Xi}_{\mathcal{D}}(x;\bm{\lambda})$
and the multi-indexed orthogonal polynomial
$\check{P}_{\mathcal{D},n}(x;\bm{\lambda})$ ($n\in\mathbb{Z}_{\geq 0}$) as
follows:
\begin{align}
  \check{\Xi}_{\mathcal{D}}(x;\bm{\lambda})&\eqdef
  \mathcal{C}_{\mathcal{D}}(\bm{\lambda})^{-1}\varphi^{(-)}_M(x)^{-1}
  \text{W}^{(-)}_{\text{C}}
  [\check{\xi}_{d_1},\ldots,\check{\xi}_{d_M}](x;\bm{\lambda}),
  \label{XiDdef}\\
  \check{P}_{\mathcal{D},n}(x;\bm{\lambda})&\eqdef
  \mathcal{C}_{\mathcal{D},n}(\bm{\lambda})^{-1}
  \varphi^{(-)}_{M+1}(x)^{-1}
  \nu(x;\bm{\lambda}+M\tilde{\bm{\delta}})^{-1}
  \text{W}^{(-)}_{\text{C}}
  [\check{\xi}_{d_1},\ldots,\check{\xi}_{d_M},\nu\check{P}_n](x;\bm{\lambda})
  \label{PDndef}\\
  &=\mathcal{C}_{\mathcal{D},n}(\bm{\lambda})^{-1}
  \varphi^{(-)}_{M+1}(x)^{-1}\n
  &\quad\times\left|
  \begin{array}{cccc}
  \check{\xi}_{d_1}(x_1)&\cdots&\check{\xi}_{d_M}(x_1)
  &r_1(x_1)\check{P}_n(x_1)\\
  \check{\xi}_{d_1}(x_2)&\cdots&\check{\xi}_{d_M}(x_2)
  &r_2(x_2)\check{P}_n(x_2)\\
  \vdots&\cdots&\vdots&\vdots\\
  \check{\xi}_{d_1}(x_{M+1})&\cdots&\check{\xi}_{d_M}(x_{M+1})
  &r_{M+1}(x_{M+1})\check{P}_n(x_{M+1})\\
  \end{array}\right|,
  \label{PDn}
\end{align}
where $x_j\eqdef x-j+1$ and $r_j(x)=r_j(x;\bm{\lambda},M)$ ($1\leq j\leq M+1$)
are given by
\begin{equation}
  r_j\bigl(x-j+1;\bm{\lambda},M\bigr)\eqdef
  \frac{\nu(x-j+1;\bm{\lambda})}{\nu(x;\bm{\lambda}+M\tilde{\bm{\delta}})}
  =\frac{(bq^{-M+x};q)_{M-j+1}(q^{x-j+2};q)_{j-1}}{(bq^{-M};q)_M},
  \label{rj}
\end{equation}
and the constants $\mathcal{C}_{\mathcal{D}}(\bm{\lambda})$ and
$\mathcal{C}_{\mathcal{D},n}(\bm{\lambda})$ are given by
\begin{align}
  \mathcal{C}_{\mathcal{D}}(\bm{\lambda})&\eqdef
  \frac{1}{\varphi^{(-)}_M(-1)}
  \prod_{1\leq j<k\leq M}\frac{\tilde{\mathcal{E}}_{d_j}(\bm{\lambda})
  -\tilde{\mathcal{E}}_{d_k}(\bm{\lambda})}
  {\alpha(\bm{\lambda})D'(-j;\bm{\lambda})},
  \label{CD}\\
  \mathcal{C}_{\mathcal{D},n}(\bm{\lambda})&\eqdef
  (-1)^Mq^{\binom{M+1}{2}}\mathcal{C}_{\mathcal{D}}(\bm{\lambda}).
  \label{CDn}
\end{align}
They are polynomials in $\eta(x)$,
\begin{alignat}{2}
  \check{\Xi}_{\mathcal{D}}(x;\bm{\lambda})&\eqdef
  \check{\Xi}_{\mathcal{D}}\bigl(\eta(x);\bm{\lambda}\bigr),&\quad
  \deg\Xi_{\mathcal{D}}(\eta)&=\ell_{\mathcal{D}},
  \label{XiDdeg}\\
  \check{P}_{\mathcal{D},n}(x;\bm{\lambda})&\eqdef
  \check{P}_{\mathcal{D},n}\bigl(\eta(x);\bm{\lambda}\bigr),&\quad
  \deg P_{\mathcal{D},n}(\eta)&=\ell_{\mathcal{D}}+n,
  \label{PDndeg}
\end{alignat}
where $\ell_{\mathcal{D}}$ is
\begin{equation}
  \ell_{\mathcal{D}}\eqdef\sum_{j=1}^Md_j-\frac12M(M-1).
  \label{lD}
\end{equation}
Their normalizations are
\begin{equation}
  \check{\Xi}_{\mathcal{D}}(-1;\bm{\lambda})=1,\quad
  \check{P}_{\mathcal{D},n}(0;\bm{\lambda})=1,
  \label{XiD(-1)PDn(0)}
\end{equation}
and we set $\check{P}_{\mathcal{D},n}(x)=0$ ($n\in\mathbb{Z}_{<0}$).
The coefficients of the highest degree terms are
\begin{align}
  \check{\Xi}_{\mathcal{D}}(x;\bm{\lambda})
  &=c^{\Xi}_{\mathcal{D}}(\bm{\lambda})\eta(x)^{\ell_{\mathcal{D}}}
  +\text{lower degree terms},\n
  c^{\Xi}_{\mathcal{D}}(\bm{\lambda})
  &=\prod_{j=1}^M\frac{\tilde{c}_{d_j}(\bm{\lambda})}
  {\tilde{c}_{j-1}(\bm{\lambda})}\cdot\!\!
  \prod_{1\leq j<k\leq M}\frac{bq^{-1}-aq^{j-1+k-1}}{bq^{-1}-aq^{d_j+d_k}},
  \label{cXiDdef}\\
  \check{P}_{\mathcal{D},n}(x;\bm{\lambda})
  &=c^P_{\mathcal{D},n}(\bm{\lambda})\eta(x)^{\ell_{\mathcal{D}}+n}
  +\text{lower degree terms},\n
  c^P_{\mathcal{D},n}(\bm{\lambda})
  &=c^{\Xi}_{\mathcal{D}}(\bm{\lambda})c_n(\bm{\lambda})
  q^{-nM}\prod_{j=1}^M\frac{1-bq^{n-d_j-1}}{1-bq^{-j}}.
  \label{cPDndef}
\end{align}
The denominator polynomial $\check{\Xi}_{\mathcal{D}}(x)$ is positive for
$x\in\mathbb{Z}_{\geq -1}$ (see the remark below \eqref{phiDnII}).
The multi-indexed orthogonal polynomial $P_{\mathcal{D},n}(\eta)$
has $n$ zeros in the physical region $0\leq\eta<1$
($\Leftrightarrow x\in\mathbb{R}_{\geq 0}$),
which interlace the $n+1$ zeros of $P_{\mathcal{D},n+1}(\eta)$ in the physical
region, and $\ell_{\mathcal{D}}$ zeros in the unphysical region
$\eta\in\mathbb{C}\backslash[0,1)$.
This property can be verified by numerical calculation.
The lowest degree multi-indexed orthogonal polynomial is related to the
denominator polynomial as
\begin{equation}
  \check{P}_{\mathcal{D},0}(x;\bm{\lambda})
  =\check{\Xi}_{\mathcal{D}}(x-1;\bm{\lambda}+\bm{\delta}).
  \label{PD0=XiD}
\end{equation}

The deformed potential functions $B_{\mathcal{D}}(x)$ \eqref{BDII} and
$D_{\mathcal{D}}(x)$ \eqref{DDII} and the eigenvectors
$\phi_{\mathcal{D}\,n}^{\text{gen}}(x)$ \eqref{phiDnII} become
\begin{align}
  B_{\mathcal{D}}(x;\bm{\lambda})&=B(x;\bm{\lambda}+M\tilde{\bm{\delta}})
  \frac{\check{\Xi}_{\mathcal{D}}(x-1;\bm{\lambda})}
  {\check{\Xi}_{\mathcal{D}}(x;\bm{\lambda})}
  \frac{\check{\Xi}_{\mathcal{D}}(x;\bm{\lambda}+\bm{\delta})}
  {\check{\Xi}_{\mathcal{D}}(x-1;\bm{\lambda}+\bm{\delta})},
  \label{BDIIJ}\\
  D_{\mathcal{D}}(x;\bm{\lambda})&=D(x)
  \frac{\check{\Xi}_{\mathcal{D}}(x;\bm{\lambda})}
  {\check{\Xi}_{\mathcal{D}}(x-1;\bm{\lambda})}
  \frac{\check{\Xi}_{\mathcal{D}}(x-2;\bm{\lambda}+\bm{\delta})}
  {\check{\Xi}_{\mathcal{D}}(x-1;\bm{\lambda}+\bm{\delta})},
  \label{DDIIJ}\\
  \phi_{\mathcal{D}\,n}^{\text{gen}}(x;\bm{\lambda})
  &=\sqrt{(bq^{-M};q)_M}\,\frac{\phi_0(x;\bm{\lambda}+M\tilde{\bm{\delta}})}
  {\sqrt{\check{\Xi}_{\mathcal{D}}(x;\bm{\lambda})
  \check{\Xi}_{\mathcal{D}}(x-1;\bm{\lambda})}}
  \check{P}_{\mathcal{D},n}(x;\bm{\lambda}).
  \label{phiDnIIJ}
\end{align}
We define the eigenvectors $\phi_{\mathcal{D}\,n}(x)$ as
\begin{align}
  \phi_{\mathcal{D}\,n}(x;\bm{\lambda})&\eqdef
  \psi_{\mathcal{D}}(x;\bm{\lambda})\check{P}_{\mathcal{D},n}(x;\bm{\lambda}),
  \quad\phi_{\mathcal{D}\,n}(0;\bm{\lambda})=1,
  \label{phiDndef}\\
  \psi_{\mathcal{D}}(x;\bm{\lambda})&\eqdef
  \sqrt{\check{\Xi}_{\mathcal{D}}(0;\bm{\lambda})}\,
  \frac{\phi_0(x;\bm{\lambda}+M\tilde{\bm{\delta}})}
  {\sqrt{\check{\Xi}_{\mathcal{D}}(x;\bm{\lambda})
  \check{\Xi}_{\mathcal{D}}(x-1;\bm{\lambda})}},\quad
  \psi_{\mathcal{D}}(0;\bm{\lambda})=1.
  \label{psiDdef}
\end{align}
Note that the formula \eqref{phi0} gives
\begin{equation*}
  \phi_{\mathcal{D}\,0}(x;\bm{\lambda})
  =\sqrt{\prod_{y=0}^{x-1}\frac{B_{\mathcal{D}}(y;\bm{\lambda})}
  {D_{\mathcal{D}}(y+1;\bm{\lambda})}}
  =\sqrt{\check{\Xi}_{\mathcal{D}}(0;\bm{\lambda})}\,
  \frac{\phi_0(x;\bm{\lambda}+M\tilde{\bm{\delta}})}
  {\sqrt{\check{\Xi}_{\mathcal{D}}(x;\bm{\lambda})
  \check{\Xi}_{\mathcal{D}}(x-1;\bm{\lambda})}}
  \check{P}_{\mathcal{D},0}(x;\bm{\lambda}),
\end{equation*}
where \eqref{XiD(-1)PDn(0)} and \eqref{PD0=XiD} are used.
The orthogonality relations for $\phi_{\mathcal{D}\,n}^{\text{gen}}(x)$
\eqref{orthophiDn} (with $N\to\infty$) give those for
$\check{P}_{\mathcal{D},n}(x)$,
\begin{equation}
  \sum_{x=0}^{\infty}\frac{\phi_0(x;\bm{\lambda}+M\tilde{\bm{\delta}})^2}
  {\check{\Xi}_{\mathcal{D}}(x;\bm{\lambda})
  \check{\Xi}_{\mathcal{D}}(x-1;\bm{\lambda})}
  \check{P}_{\mathcal{D},n}(x;\bm{\lambda})
  \check{P}_{\mathcal{D},m}(x;\bm{\lambda})
  =\frac{\delta_{nm}}
  {d_n(\bm{\lambda})^2\tilde{d}_{\mathcal{D},n}(\bm{\lambda})^2}
  \ \ (n,m\in\mathbb{Z}_{\geq 0}),
  \label{orthoPDn}
\end{equation}
where $\tilde{d}_{\mathcal{D},n}(\bm{\lambda})$ ($>0$) is given by
\begin{equation}
  \tilde{d}_{\mathcal{D},n}(\bm{\lambda})^2
  \eqdef\kappa^{\binom{M}{2}}\prod_{j=1}^M
  \frac{\alpha(\bm{\lambda})D'(0;\bm{\lambda}+(j-1)\tilde{\bm{\delta}})}
  {\mathcal{E}_n(\bm{\lambda})-\tilde{\mathcal{E}}_{d_j}(\bm{\lambda})}
  =\frac{(bq^{-M};q)_M}{\prod_{j=1}^M
  \bigl(\mathcal{E}_n(\bm{\lambda})-\tilde{\mathcal{E}}_{d_j}(\bm{\lambda})
  \bigr)}.
\end{equation}

The Hamiltonian of the deformed system is \eqref{HD} (with $N\to\infty$),
\begin{align}
  &\quad\mathcal{H}_{\mathcal{D}}(\bm{\lambda})
  =\bigl(\mathcal{H}_{\mathcal{D}}(\bm{\lambda})_{x,y}
  \bigr)_{x,y\in\mathbb{Z}_{\geq 0}}
  =\mathcal{A}_{\mathcal{D}}(\bm{\lambda})^{\dagger}
  \mathcal{A}_{\mathcal{D}}(\bm{\lambda})
  \label{HDJ}\\
  &\eqdef-\sqrt{B_{\mathcal{D}}(x;\bm{\lambda})}\,e^{\partial}
  \sqrt{D_{\mathcal{D}}(x;\bm{\lambda})}
  -\sqrt{D_{\mathcal{D}}(x;\bm{\lambda})}\,e^{-\partial}
  \sqrt{B_{\mathcal{D}}(x;\bm{\lambda})}
  +B_{\mathcal{D}}(x;\bm{\lambda})+D_{\mathcal{D}}(x;\bm{\lambda}).\nonumber
\end{align}
The eigenvectors $\phi_{\mathcal{D}\,n}(x)$ \eqref{phiDndef} satisfy the
Schr\"odinger equation,
\begin{equation}
  \mathcal{H}_{\mathcal{D}}(\bm{\lambda})\phi_{\mathcal{D}\,n}(x;\bm{\lambda})
  =\mathcal{E}_n(\bm{\lambda})\phi_{\mathcal{D}\,n}(x;\bm{\lambda})
  \ \ (n\in\mathbb{Z}_{\geq 0}).
\end{equation}
The similarity transformed Hamiltonian is defined by
\begin{align}
  \widetilde{\mathcal{H}}_{\mathcal{D}}(\bm{\lambda})
  &\eqdef\psi_{\mathcal{D}}(x;\bm{\lambda})^{-1}\circ
  \mathcal{H}_{\mathcal{D}}(\bm{\lambda})\circ
  \psi_{\mathcal{D}}(x;\bm{\lambda})\n
  &=B(x;\bm{\lambda}+M\tilde{\bm{\delta}})\,
  \frac{\check{\Xi}_{\mathcal{D}}(x-1;\bm{\lambda})}
  {\check{\Xi}_{\mathcal{D}}(x;\bm{\lambda})}
  \biggl(\frac{\check{\Xi}_{\mathcal{D}}(x;\bm{\lambda}+\bm{\delta})}
  {\check{\Xi}_{\mathcal{D}}(x-1;\bm{\lambda}+\bm{\delta})}-e^{\partial}
  \biggr)\n
  &\quad+D(x)\,
  \frac{\check{\Xi}_{\mathcal{D}}(x;\bm{\lambda})}
  {\check{\Xi}_{\mathcal{D}}(x-1;\bm{\lambda})}
  \biggl(\frac{\check{\Xi}_{\mathcal{D}}(x-2;\bm{\lambda}+\bm{\delta})}
  {\check{\Xi}_{\mathcal{D}}(x-1;\bm{\lambda}+\bm{\delta})}-e^{-\partial}
  \biggr),
  \label{HDtdef}
\end{align}
and the multi-indexed orthogonal polynomials
$\check{P}_{\mathcal{D},n}(x;\bm{\lambda})$ are its eigenpolynomials:
\begin{equation}
  \widetilde{\mathcal{H}}_{\mathcal{D}}(\bm{\lambda})
  \check{P}_{\mathcal{D},n}(x;\bm{\lambda})=\mathcal{E}_n(\bm{\lambda})
  \check{P}_{\mathcal{D},n}(x;\bm{\lambda})\ \ (n\in\mathbb{Z}_{\geq 0}).
  \label{HDtPDn=}
\end{equation}

The shape invariance of the original system \eqref{si} is inherited by
the deformed systems:
\begin{equation}
  \mathcal{A}_{\mathcal{D}}(\bm{\lambda})
  \mathcal{A}_{\mathcal{D}}(\bm{\lambda})^{\dagger}
  =\kappa\mathcal{A}_{\mathcal{D}}(\bm{\lambda}+\bm{\delta})^{\dagger}
  \mathcal{A}_{\mathcal{D}}(\bm{\lambda}+\bm{\delta})
  +\mathcal{E}_1(\bm{\lambda}).
  \label{shapeinvD}
\end{equation}
As a consequence of the shape invariance and the normalization,
we obtain
\begin{align}
  &\mathcal{A}_{\mathcal{D}}(\bm{\lambda})
  \phi_{\mathcal{D}\,n}(x;\bm{\lambda})
  =\frac{\mathcal{E}_n(\bm{\lambda})}
  {\sqrt{B_{\mathcal{D}}(0;\bm{\lambda})}}\,
  \phi_{\mathcal{D}\,n-1}(x;\bm{\lambda}+\bm{\delta})
  \ \ (n\in\mathbb{Z}_{\geq 0}),
  \label{ADphiDn=}\\
  &\mathcal{A}_{\mathcal{D}}(\bm{\lambda})^{\dagger}
  \phi_{\mathcal{D}\,n-1}(x;\bm{\lambda}+\bm{\delta})
  =\sqrt{B_{\mathcal{D}}(0;\bm{\lambda})}\,
  \phi_{\mathcal{D}\,n}(x;\bm{\lambda})
  \ \ (n\in\mathbb{Z}_{\geq 1}).
  \label{ADdphiDn-1=}
\end{align}
These relations give the forward and backward shift relations:
\begin{align}
  &\mathcal{F}_{\mathcal{D}}(\bm{\lambda})
  \check{P}_{\mathcal{D},n}(x;\bm{\lambda})
  =\mathcal{E}_n(\bm{\lambda})
  \check{P}_{\mathcal{D},n-1}(x;\bm{\lambda}+\bm{\delta})
  \ \ (n\in\mathbb{Z}_{\geq 0}),
  \label{FDPDn=}\\
  &\mathcal{B}_{\mathcal{D}}(\bm{\lambda})
  \check{P}_{\mathcal{D},n-1}(x;\bm{\lambda}+\bm{\delta})
  =\check{P}_{\mathcal{D},n}(x;\bm{\lambda})
  \ \ (n\in\mathbb{Z}_{\geq 1}),
  \label{BDPDn-1=}
\end{align}
where the forward and backward shift operators are
\begin{align}
  \mathcal{F}_{\mathcal{D}}(\bm{\lambda})&\eqdef
  \sqrt{B_{\mathcal{D}}(0;\bm{\lambda})}\,
  \psi_{\mathcal{D}}(x;\bm{\lambda}+\bm{\delta})^{-1}\circ
  \mathcal{A}_{\mathcal{D}}(\bm{\lambda})\circ
  \psi_{\mathcal{D}}(x;\bm{\lambda})\n
  &=\frac{B(0;\bm{\lambda}+M\tilde{\bm{\delta}})}
  {\varphi(x)\check{\Xi}_{\mathcal{D}}(x;\bm{\lambda})}
  \Bigl(\check{\Xi}_{\mathcal{D}}(x;\bm{\lambda}+\bm{\delta})
  -\check{\Xi}_{\mathcal{D}}(x-1;\bm{\lambda}+\bm{\delta})e^{\partial}\Bigr),
  \label{calFD}\\
  \mathcal{B}_{\mathcal{D}}(\bm{\lambda})&\eqdef
  \frac{1}{\sqrt{B_{\mathcal{D}}(0;\bm{\lambda})}}\,
  \psi_{\mathcal{D}}(x;\bm{\lambda})^{-1}\circ
  \mathcal{A}_{\mathcal{D}}(\bm{\lambda})^{\dagger}\circ
  \psi_{\mathcal{D}}(x;\bm{\lambda}+\bm{\delta})\n
  &=\frac{1}{B(0;\bm{\lambda}+M\tilde{\bm{\delta}})
  \check{\Xi}_{\mathcal{D}}(x-1;\bm{\lambda}+\bm{\delta})}\n
  &\quad\times
  \Bigl(B(x;\bm{\lambda}+M\tilde{\bm{\delta}})
  \check{\Xi}_{\mathcal{D}}(x-1;\bm{\lambda})
  -D(x)\check{\Xi}_{\mathcal{D}}(x;\bm{\lambda})e^{-\partial}\Bigr)
  \varphi(x).
  \label{calBD}
\end{align}
The similarity transformed Hamiltonian \eqref{HDtdef} is expressed as
$\widetilde{\mathcal{H}}_{\mathcal{D}}(\bm{\lambda})
=\mathcal{B}_{\mathcal{D}}(\bm{\lambda})\mathcal{F}_{\mathcal{D}}(\bm{\lambda})$.

The denominator polynomial $\check{\Xi}_{\mathcal{D}}(x)$ and the multi-indexed
polynomials $\check{P}_{\mathcal{D},n}(x)$ are normalized as
\eqref{XiD(-1)PDn(0)}.
Their values at $x=\infty$ are given by (cf. \eqref{Pn(0)} and \eqref{xiv(-1)})
\begin{align}
  \check{\Xi}_{\mathcal{D}}(\infty;\bm{\lambda})\bigl(\,\eqdef
  \lim_{x\to\infty}\check{\Xi}_{\mathcal{D}}(x;\bm{\lambda})\bigr)
  &=\prod_{j=1}^M\frac{\tilde{c}'_{d_j}(\bm{\lambda})}
  {\tilde{c}'_{j-1}(\bm{\lambda})},
  \label{XiDinf}\\
  \check{P}_{\mathcal{D},n}(\infty;\bm{\lambda})\bigl(\,\eqdef
  \lim_{x\to\infty}\check{P}_{\mathcal{D},n}(x;\bm{\lambda})\bigr)
  &=\prod_{j=1}^M\frac{\tilde{c}'_{d_j}(\bm{\lambda})}
  {\tilde{c}'_{j-1}(\bm{\lambda})}\cdot
  \prod_{j=1}^M\frac{\mathcal{E}_n(\bm{\lambda})
  -\tilde{\mathcal{E}}_{d_j}(\bm{\lambda})}
  {-\tilde{\mathcal{E}}_{j-1}(\bm{\lambda})}
  \cdot c'_n(\bm{\lambda}).
  \label{PDninf}
\end{align}

In \eqref{setD}, we have assumed the order $d_1<d_2<\cdots<d_M$ (standard order).
Under the permutations of $d_j$'s, the denominator polynomial
$\check{\Xi}_{\mathcal{D}}(x)$ \eqref{XiDdef} and the multi-indexed polynomial
$\check{P}_{\mathcal{D},n}(x)$ \eqref{PDndef} may change their sign, but the
deformed potential functions $B_{\mathcal{D}}(x)$ \eqref{BDIIJ} and
$D_{\mathcal{D}}(x)$ \eqref{DDIIJ} are invariant. So the deformed Hamiltonian
$\mathcal{H}_{\mathcal{D}}$ \eqref{HDJ} does not depend on the order of $d_j$'s.
Setting one of $d_j$ to $0$, for example $d_M=0$, we obtain the following
relation between $M$-indexed polynomial and $(M-1)$-indexed polynomial,
\begin{equation}
  \check{P}_{\mathcal{D},n}(x;\bm{\lambda})\Bigm|_{d_M=0}
  =\check{P}_{\mathcal{D}',n}(x;\bm{\lambda}+\tilde{\bm{\delta}}),\quad
  \mathcal{D}'=\{d_1-1,d_2-1,\ldots,d_{M-1}-1\}.
  \label{dM=0}
\end{equation}
The denominator polynomial $\check{\Xi}_{\mathcal{D}}(x)$ behaves similarly.
This is why we have restricted $d_j\geq 1$.

We have assumed the parameter range \eqref{range}.
This range is needed in the intermediate Darboux transformations, but may be
extended in the final results of the deformed system.
There is one more reason why the range may be extended.
Following our previous papers \cite{os26,os35}, we have used the symmetry
\eqref{H=aH'+a'} and imposed positivity on $B'(x)$, $D'(x)$ and $\alpha$,
respectively.
However, $B'(x)$ and $D'(x)$ always appear in combination with $\alpha$.
So, by introducing the following new quantities,
\begin{align}
  &B^{\prime\,\text{new}}(x)\eqdef\alpha B'(x),\quad
  D^{\prime\,\text{new}}(x)\eqdef\alpha D'(x),\quad
  \mathcal{E}^{\prime\,\text{new}}_{\text{v}}\eqdef
  \alpha\mathcal{E}'_{\text{v}},
  \label{B'new}\\
  &\mathcal{H}^{\prime\,\text{new}}\eqdef
  -\sqrt{B^{\prime\,\text{new}}(x)}\,e^{\partial}
  \sqrt{D^{\prime\,\text{new}}(x)}
  -\sqrt{D^{\prime\,\text{new}}(x)}\,e^{-\partial}
  \sqrt{B^{\prime\,\text{new}}(x)}
  +B^{\prime\,\text{new}}(x)+D^{\prime\,\text{new}}(x),\nonumber
\end{align}
the symmetry \eqref{H=aH'+a'} is expressed as
$\mathcal{H}=\mathcal{H}^{\prime\,\text{new}}+\alpha'$ and the positivity
condition is imposed on $B^{\prime\,\text{new}}(x)$ and
$D^{\prime\,\text{new}}(x)$.
The virtual state energy \eqref{Etv<0} is $\tilde{\mathcal{E}}_{\text{v}}=
\mathcal{E}^{\prime\,\text{new}}_{\text{v}}+\alpha'$
and it should be negative for $\text{v}\in\mathcal{D}$ and $\text{v}=0$
($\alpha'=\tilde{\mathcal{E}}_0$).
By considering these, the parameter range is extended to
\begin{equation}
  0<a<1,\quad b<q^{1+d_M}.
  \label{range_ext}
\end{equation}

\subsection{Examples}
\label{sec:Ex}

For illustration, let us write down the type $\I$ and $\II$ single-indexed
little $q$-Jacobi polynomials.
For $\mathcal{D}=\{d\}$, they are given by
\begin{align}
  \check{P}^{\I}_{\{d\},n}(x)&=
  \frac{(1-b)q^n}{(1-aq^{n-d-1})(1-bq^{n+d})}\bigl(
  \check{\xi}^{\I}_d(x+1)\check{P}_n(x)
  -aq^{-1}\check{\xi}^{\I}_d(x)\check{P}_n(x+1)\bigr),\\
  \check{P}^{\II}_{\{d\},n}(x)&=
  \frac{q^{-x}}{1-bq^{-1}}\bigl(
  (1-bq^{x-1})\check{\xi}^{\II}_d(x-1)\check{P}_n(x)
  -(1-q^x)\check{\xi}^{\II}_d(x)\check{P}_n(x-1)\bigr).
\end{align}
The type $\I$ and $\II$ multi-indexed polynomials are different polynomials,
except for $\mathcal{D}=\{1\}$, in which case the two polynomials are related as
$\check{P}^{\I}_{\{1\},n}(x;\bm{\lambda}-\tilde{\bm{\delta}}^{\I})
=\check{P}^{\II}_{\{1\},n}(x;\bm{\lambda}-\tilde{\bm{\delta}}^{\II})$.
For example, for $\mathcal{D}=\{2\}$ and $n=0,1$, we have
\begin{align*}
  \check{P}^{\I}_{\{2\},0}(x)&=\frac{1}{(1-bq)(1-bq^2)}
  \bigl((1-aq^{-1})(1-aq^{-2})+(1+q)(1-aq^{-2})(a-bq^3)q^{x-2}\\
  &\qquad+(a-bq^3)(a-bq^4)q^{2x-4}\bigr),\\
  \check{P}^{\I}_{\{2\},1}(x)&=\frac{q}{a(1-b)(1-bq)(1-bq^3)}
  \Bigl(-(1-a)(1-aq^{-1})(1-aq^{-3})\\
  &\qquad-(1-a)(1-aq^{-3})\bigl((1+q)(a-bq^3)-q^2(1-ab)\bigr)q^{x-2}\\
  &\qquad+(1-aq^{-3})(a-bq^3)\bigl((1+q)(1-abq)-a+bq^2\bigr)q^{2x-2}\\
  &\qquad+(1-ab)(a-bq^3)(a-bq^4)q^{3x-4}\Bigr),\\
  \check{P}^{\II}_{\{2\},0}(x)&=\frac{1}{(1-bq^{-1})(1-bq^{-2})}
  \bigl((1-aq)(1-aq^2)-(1+q)(1-aq^2)(b-aq^3)q^{x-2}\\
  &\qquad+(b-aq^3)(b-aq^4)q^{2x-3}\bigr),\\
  \check{P}^{\II}_{\{2\},1}(x)&=\frac{q^{-1}}{a(1-b)(1-bq^{-1})(1-bq^{-3})}
  \Bigl(-(1-a)(1-aq)(1-aq^3)\\
  &\qquad+(1-a)(1-aq^3)\bigl((1+q)(b-aq^3)+q^2(1-ab)\bigr)q^{x-2}\\
  &\qquad-(1-aq^3)(b-aq^3)\bigl((1+q)(q-ab)+b-aq^2\bigr)q^{2x-3}\\
  &\qquad+(1-ab)(b-aq^3)(b-aq^4)q^{3x-3}\Bigr).
\end{align*}
We remark that
$\check{P}^{\I}_{\{2\},n}(x;\bm{\lambda})\bigl|_{x\to-x,q\to q^{-1}}
=\check{P}^{\II}_{\{2\},n}(x;\bm{\lambda})$ holds for $n=0,1$,
but does not hold for $n\geq 2$.

\section{Multi-indexed Little $q$-Laguerre polynomials}
\label{sec:mioplqL}

In this section we present the case-(1) multi-indexed little $q$-Laguerre
polynomials, especially type $\II$ polynomials.

\subsection{Original system}
\label{sec:lqL}

Let us present the basic data of little $q$-Laguerre rdQM system.
The standard little $q$-Laguerre polynomial
$p_n(q^x;a|q)={}_2\phi_1\bigl(\genfrac{}{}{0pt}{}{q^{-n},\,0}{aq}
\bigl|q\,;q^{x+1}\bigr)$ \cite{kls} does not satisfy our normalization
$\check{P}_n(0)=1$.
The little $q$-Laguerre system is obtained from little $q$-Jacobi system by
setting $b=0$.
We change the parametrization slightly from the standard one,
$a^{\text{standard}}=aq^{-1}$.
The basic data are as follows \cite{os12,os34}:
\begin{align}
  &q^{\bm{\lambda}}=a,\quad
  \bm{\delta}=1,\quad \kappa=q^{-1},\quad 0<a<1,\\
  &B(x;\bm{\lambda})=aq^{-x-1},\quad D(x)=q^{-x}-1,\\
  &\mathcal{E}_n=q^{-n}-1,\quad
  \eta(x)=1-q^x,\quad\varphi(x)=q^x,\\
  &\check{P}_n(x;\bm{\lambda})
  ={}_2\phi_0\Bigl(
  \genfrac{}{}{0pt}{}{q^{-n},\,q^{-x}}{-}\Bigm|
  q\,;a^{-1}q^{x+1}\Bigr)
  =c'_n(\bm{\lambda})\,p_n\bigl(1-\eta(x);aq^{-1}|q\bigr)\n
  &\phantom{\check{P}_n(x;\bm{\lambda})}
  =c'_n(\bm{\lambda})\,{}_2\phi_1\Bigl(
  \genfrac{}{}{0pt}{}{q^{-n},\,0}{a}\Bigm|q\,;q^{x+1}\Bigr),\quad
  c'_n(\bm{\lambda})\eqdef(-a)^{-n}q^{-\binom{n}{2}}(a;q)_n,\\
  &\check{P}_n(x;\bm{\lambda})=c_n(\bm{\lambda})\eta(x)^n
  +\text{lower degree terms},
  \quad c_n(\bm{\lambda})\eqdef(-a)^{-n}q^{-n(n-1)},\\
  &\phi_0(x;\bm{\lambda})^2=\frac{a^x}{(q;q)_x}
  =\frac{(q^x;q)_{\infty}}{(q;q)_{\infty}}a^x,\quad
  \phi_0(x;\bm{\lambda})>0,\quad\phi_0(0;\bm{\lambda})=1,
  \label{lqLphi0}\\
  &d_n(\bm{\lambda})^2
  =\frac{a^nq^{n(n-1)}}{(a,q;q)_n}\times(a;q)_{\infty},\quad
  d_n(\bm{\lambda})>0,
\end{align}
and $\check{P}_n(0;\bm{\lambda})$ satisfies \eqref{Pn(0)}.
The little $q$-Laguerre system has shape invariance \eqref{si} and the formulas
\eqref{Aphin=}--\eqref{varphiM} hold.

\subsection{Type $\I$ polynomials}
\label{sec:lqLI}

The potential functions $B'(x)$ and $D'(x)$ and the virtual state polynomials
$\check{\xi}_{\text{v}}(x)$ are given by \eqref{B'I}--\eqref{xiI},
where the twist operation $\mathfrak{t}$ and the shift $\tilde{\bm{\delta}}$ are
(remark: $a^{\text{standard}}$ is used in \cite{os35})
\begin{align}
  &\mathfrak{t}^{\I}(\bm{\lambda})\eqdef 2-\lambda_1,
  \ \ \text{namely}\ \ q^{\mathfrak{t}^{\I}(\bm{\lambda})}=a^{-1}q^2,\\
  &\tilde{\bm{\delta}}^{\I}\eqdef -1,\quad
  \mathfrak{t}^{\I}(\bm{\lambda})+u\bm{\delta}=
  \mathfrak{t}^{\I}(\bm{\lambda}+u\tilde{\bm{\delta}}^{\I})
  \ \ (\forall u\in\mathbb{R}),
\end{align}
with $\alpha^{\I}(\bm{\lambda})\eqdef aq^{-1}$ and
$\alpha^{\prime\,\I}(\bm{\lambda})\eqdef-(1-aq^{-1})$.
The parameter range is $0<a<q^{1+d_M}$.
Various formulas for the type $\I$ multi-indexed little $q$-Laguerre polynomials
are presented in \cite{os35}, which are obtained from those for the type $\I$
multi-indexed little $q$-Jacobi polynomials by setting $b=0$.

\subsection{Type $\II$ polynomials}
\label{sec:lqLII}

The potential functions $B'(x)$ and $D'(x)$ for little $q$-Jacobi system
\eqref{B'II} diverge in the $b\to 0$ limit.
But $B^{\prime\,\text{new}}(x)$ and $D^{\prime\,\text{new}}(x)$ \eqref{B'new}
have well-defined $b\to 0$ limits, because $\alpha$ \eqref{alphaII} vanishes.
So we define $B^{\prime\,\text{new}}(x)$ and $D^{\prime\,\text{new}}(x)$ as
follows:
\begin{equation}
  B^{\prime\,\text{new}\,\II}(x;\bm{\lambda})\eqdef a(q^{-x-1}-1),\quad
  D^{\prime\,\text{new}\,\II}(x)\eqdef q^{-x}.
\end{equation}
The constant $\alpha'$ and the shift $\tilde{\bm{\delta}}$ are defined by
\begin{equation}
  \alpha^{\prime\,\II}(\bm{\lambda})\eqdef-(1-a),\quad
  \tilde{\bm{\delta}}^{\II}\eqdef 1.
\end{equation}
By taking $b\to 0$ limit of \eqref{xivII}, the virtual state polynomial
$\check{\xi}_{\text{v}}(x)$ ($\text{v}\in\mathbb{Z}_{\geq 0}$) is given by
\begin{equation}
  \check{\xi}^{\II}_{\text{v}}(x;\bm{\lambda})\eqdef
  \tilde{c}^{\prime\,\II}_{\text{v}}(\bm{\lambda})\,{}_1\phi_1\Bigl(
  \genfrac{}{}{0pt}{}{q^{-\text{v}}}{a}
  \Bigm|q\,;aq^{x+\text{v}+1}\Bigr),\quad
  \tilde{c}^{\prime\,\II}_{\text{v}}(\bm{\lambda})\eqdef(a;q)_{\text{v}},
\end{equation}
which satisfies \eqref{xiv(-1)} and \eqref{ctvII} with
\begin{equation}
  \tilde{c}^{\II}_{\text{v}}(\bm{\lambda})\eqdef
  (-a)^{\text{v}}q^{\text{v}^2}.
\end{equation}
For simplicity of presentation, the superscript $\II$ is omitted in the
following.

The virtual state polynomial $\check{\xi}_{\text{v}}(x)$ satisfies the
difference equation (for $x\in\mathbb{R}$) \eqref{xieq} (with superscript
``new'') with
\begin{equation}
  \mathcal{E}^{\prime\,\text{new}}_{\text{v}}(\bm{\lambda})\eqdef
  -a(1-q^{\text{v}}).
\end{equation}
It is positive $\check{\xi}_{\text{v}}(x)>0$ ($x\in\mathbb{Z}_{\geq-1}$) for
$0<a<1$.

In the following, we assume the following parameter range:
\begin{equation}
  0<a<1.
  \label{rangeL}
\end{equation}
The functions $\tilde{\phi}_0(x)$ ($>0$) \eqref{phit0} and $\nu(x)$ \eqref{nu}
become
\begin{align}
  \tilde{\phi}_0(x;\bm{\lambda})^2&=(q;q)_xa^x
  =\frac{(q;q)_{\infty}}{(q^{x+1};q)_{\infty}}a^x,\\
  \nu(x;\bm{\lambda})&=\frac{1}{(q;q)_x}
  =\frac{(q^{x+1};q)_{\infty}}{(q;q)_{\infty}},
\end{align}
and the virtual state energy $\tilde{\mathcal{E}}_{\text{v}}$ \eqref{Etv<0}
becomes
\begin{equation}
  \tilde{\mathcal{E}}_{\text{v}}(\bm{\lambda})
  =-(1-aq^{\text{v}}).
\end{equation}
The virtual state vectors $\tilde{\phi}_{\text{v}}(x)$ \eqref{phitv} satisfy
the Schr\"odinger equation except for the end-point $x=0$,
\eqref{SchphitvII} (with $N\to\infty$).

The denominator polynomial $\check{\Xi}_{\mathcal{D}}(x;\bm{\lambda})$
and the multi-indexed orthogonal polynomial
$\check{P}_{\mathcal{D},n}(x;\bm{\lambda})$ ($n\in\mathbb{Z}_{\geq 0}$) are
defined by \eqref{XiDdef}--\eqref{CDn}
($\alpha(\bm{\lambda})D'(-j;\bm{\lambda})$ is replaced by
$D^{\prime\,\text{new}}(-j;\bm{\lambda})$ in \eqref{CD}) with 
\begin{equation}
  r_j\bigl(x-j+1;\bm{\lambda},M\bigr)=(q^{x-j+2};q)_{j-1}.
\end{equation}
They are polynomials in $\eta(x)$, \eqref{XiDdeg}--\eqref{lD}, and
their normalizations are \eqref{XiD(-1)PDn(0)}.
The coefficients of the highest degree terms are
\eqref{cXiDdef}--\eqref{cPDndef} with
\begin{align}
  c^{\Xi}_{\mathcal{D}}(\bm{\lambda})
  &=\prod_{j=1}^M\frac{\tilde{c}_{d_j}(\bm{\lambda})}
  {\tilde{c}_{j-1}(\bm{\lambda})}\cdot q^{-(M-1)\ell_{\mathcal{D}}},\\
  c^P_{\mathcal{D},n}(\bm{\lambda})
  &=c^{\Xi}_{\mathcal{D}}(\bm{\lambda})c_n(\bm{\lambda})q^{-nM}.
\end{align}
The lowest degree multi-indexed orthogonal polynomial and the denominator
polynomial are related as \eqref{PD0=XiD}.
In the end, the type $\II$ multi-indexed little $q$-Laguerre polynomials are
obtained from the type $\II$ multi-indexed little $q$-Jacobi polynomials by
taking $b\to 0$ limit.

The deformed potential functions $B_{\mathcal{D}}(x)$ \eqref{BDII} and
$D_{\mathcal{D}}(x)$ \eqref{DDII} become \eqref{BDIIJ}--\eqref{DDIIJ},
and the eigenvectors $\phi_{\mathcal{D}\,n}^{\text{gen}}(x)$ \eqref{phiDnII} 
become
\begin{equation}
  \phi_{\mathcal{D}\,n}^{\text{gen}}(x;\bm{\lambda})
  =\frac{\phi_0(x;\bm{\lambda}+M\tilde{\bm{\delta}})}
  {\sqrt{\check{\Xi}_{\mathcal{D}}(x;\bm{\lambda})
  \check{\Xi}_{\mathcal{D}}(x-1;\bm{\lambda})}}
  \check{P}_{\mathcal{D},n}(x;\bm{\lambda}).
\end{equation}
We define the eigenvectors $\phi_{\mathcal{D}\,n}(x)$ as
\eqref{phiDndef}--\eqref{psiDdef}.
The orthogonality relations for $\phi_{\mathcal{D}\,n}^{\text{gen}}(x)$
\eqref{orthophiDn} (with $N\to\infty$) give those for
$\check{P}_{\mathcal{D},n}(x)$, \eqref{orthoPDn} with
\begin{equation}
  \tilde{d}_{\mathcal{D},n}(\bm{\lambda})^2
  =\frac{1}{\prod_{j=1}^M
  \bigl(\mathcal{E}_n-\tilde{\mathcal{E}}_{d_j}(\bm{\lambda})
  \bigr)}.
\end{equation}

The Hamiltonian of the deformed system, the Schr\"odinger equation and
their similarity transformed versions are given by \eqref{HDJ}--\eqref{HDtPDn=}.
The shape invariance of the original system \eqref{si} is inherited by
the deformed systems \eqref{shapeinvD}. As its consequence, we have relations
\eqref{ADphiDn=}--\eqref{ADdphiDn-1=} and the forward and backward shift
relations \eqref{FDPDn=}--\eqref{calBD}.

The denominator polynomial $\check{\Xi}_{\mathcal{D}}(x)$ and the multi-indexed
polynomials $\check{P}_{\mathcal{D},n}(x)$ are normalized as
\eqref{XiD(-1)PDn(0)}.
Their values at $x=\infty$ are given by \eqref{XiDinf}--\eqref{PDninf}.
The reason for restricting $d_j\geq 1$ is the same as for little $q$-Jacobi case,
\eqref{dM=0}.

\section{Summary and Comments}
\label{sec:summary}

We have reconsidered the multi-step Darboux transformations with the virtual
states as seed solutions for rdQM systems.
There are two types of virtual states vectors, type $\I$ and type $\II$.
For finite rdQM systems such as $q$-Racah and Racah cases, the multi-step
Darboux transformations with these two types of virtual states as seed solutions
give essentially the same multi-indexed polynomials.
On the other hand, for semi-infinite rdQM systems such as little $q$-Jacobi
and little $q$-Laguerre cases, they give different multi-indexed polynomials.
By constructing the type $\II$ virtual state vectors explicitly, we obtain the
case-(1) type $\II$ multi-indexed little $q$-Jacobi and little $q$-Laguerre
orthogonal polynomials, which satisfy second order difference equations.
The deformed rdQM systems have shape invariance and the multi-indexed polynomial
satisfy the forward and backward shift relations.
It is an interesting problem to study other semi-infinite rdQM systems such as
Meixner and $q$-Meixner cases.

In our previous studies
\cite{os25}--\cite{idQMcH}
(except for the type $\I$ Laguerre),
the virtual states are obtained from the eigenstates by twisting the parameters
(For the type $\I$ Laguerre, the virtual states are obtained from the
eigenstates by replacing $x$ with $ix$).
In the cases of the type $\II$ little $q$-Jacobi and little $q$-Laguerre,
the situation is different.
For the type $\II$ little $q$-Jacobi case in \S\,\ref{sec:lqJII}, the twist
operation is defined by \eqref{twistII}, but it is used only in \eqref{E'v}.
The potential functions $B'(x)$ and $D'(x)$ \eqref{B'II} and the virtual state
polynomial $\check{\xi}_{\text{v}}(x)$ \eqref{xivII} are not obtained from
$B(x)$, $D(x)$ and $\check{P}_n(x)$ by twisting the parameters.
This is also the case for the type $\II$ little $q$-Laguerre in
\S\,\ref{sec:lqLII}.

The little $q$-Jacobi (Laguerre) polynomials reduce to the Jacobi (Laguerre)
polynomials in the $q\to 1$ limit.
Similarly the multi-indexed little $q$-Jacobi (Laguerre) polynomials reduce to
the multi-indexed Jacobi (Laguerre) polynomials in the $q\to 1$ limit.
The type $\I$ ($\II$) multi-indexed little $q$-Jacobi polynomials
($\bm{\lambda}^{\text{l$q$J}}=(g+\frac12,h+\frac12)$) reduce to
the multi-indexed Jacobi polynomials ($\bm{\lambda}^{\text{J}}=(g,h)$) with
only type $\II$ ($\I$) indices.
For the virtual states in oQM, see \cite{os29}.
The reason for the exchange of type $\I$ and type $\II$ is the coordinate
correspondence,
$q^{x^{\text{l$q$J}}+1}\leftrightarrow\frac12(1-\cos 2x^{\text{J}})$.
The minimum and maximum values of $x^{\text{l$q$J}}$,
$x^{\text{l$q$J}}_{\text{min}}=0$ and $x^{\text{l$q$J}}_\text{max}=\infty$,
correspond to the maximum and minimum values of $x^{\text{J}}$,
$x^{\text{J}}_{\text{max}}=\frac12\pi$ and $x^{\text{J}}_{\text{min}}=0$,
respectively.
Similarly the type $\I$ ($\II$) multi-indexed little $q$-Laguerre polynomials
($\bm{\lambda}^{\text{l$q$L}}=g+\frac12$) reduce to
the multi-indexed Laguerre polynomials ($\bm{\lambda}^{\text{L}}=g$) with
only type $\II$ ($\I$) indices in the $q\to 1$ limit.
We have no twist operation for the type $\II$ little $q$-Laguerre and the type
$\I$ Laguerre.
For the multi-indexed Jacobi and Laguerre polynomials, it is possible to use
type $\I$ and $\II$ indices at the same time \cite{os25}.
It is a challenging problem to study whether mixed use of type $\I$ and $\II$
indices is possible for the multi-indexed little $q$-Jacobi and little
$q$-Laguerre polynomials.
In the type $\I$ ($\II$) construction, various quantities are expressed in
terms of the Casoratian $\text{W}_{\text{C}}$ \eqref{WC}
($\text{W}_{\text{C}}^{(-)}$ \eqref{WC-}),
in which $x$ is shifted to $+$ ($-$) direction, respectively.
Since $x$ is shifted in the opposite direction, using both $\text{W}_{\text{C}}$
and $\text{W}_{\text{C}}^{(-)}$ is not a good combination.
In the above \eqref{WC}, we mention that the type-(\romannumeral2)
(type-(\romannumeral1)) factorization for the type $\I$ (type $\II$) virtual
state vectors is impossible, respectively. But this is the case for finite $N$.
For $N\to\infty$ case, the type-(\romannumeral2) factorization for the
type $\I$ virtual state vectors is possible.
This would be a hint for mixed use of type $\I$ and $\II$ indices.


\section*{Acknowledgements}

I thank the support by Course of Physics, Department of Science.


\end{document}